\documentclass[paper,notoc]{JHEP3}
\usepackage{amssymb}   
\usepackage{amsmath}
\usepackage{graphicx}
\usepackage{axodraw4j}

\def\beq{\begin{equation}}
\def\eeq{\end{equation}}
\def\bea{\begin{eqnarray}}
\def\eea{\end{eqnarray}}

\def\r{\right}
\def\l{\left}
\def\f21{{}_2F_{1}}
\def\N{\mathcal{N}}

\def\d{ \mathrm{d}}
  
\def\A{\mathcal{A}}

\newcommand{\eps}{\epsilon}

\newcommand{\cMSOne}[0]{\mathcal{M}^S_{1}}
\newcommand{\cMSTwo}[0]{\mathcal{M}^S_{2}}
\newcommand{\cMSFour}[0]{\mathcal{M}^S_{3}}
\newcommand{\cMSSix}[0]{\mathcal{M}^S_{4}}
\newcommand{\cMSTen}[0]{\mathcal{M}^S_{5}}
\newcommand{\cMSTwelve}[0]{\mathcal{M}^S_{6}}
\newcommand{\cMSFourteen}[0]{\mathcal{M}^S_{7}}
\newcommand{\cMSFifteen}[0]{\mathcal{M}^S_{8}}
\newcommand{\cMSNineteen}[0]{\mathcal{M}^S_{9}}
\newcommand{\cMSTwentyThree}[0]{\mathcal{M}^S_{10}}
\newcommand{\cMSThirtyThree}[0]{\mathcal{M}^S_{11}}
\newcommand{\cMSThirtyFour}[0]{\mathcal{M}^S_{12}}
\newcommand{\cMSThirtySix}[0]{\mathcal{M}^S_{13}}
\newcommand{\cMSThirtySeven}[0]{\mathcal{M}^S_{14}}

\newcommand{\MSOne}[0]{{M}^S_{1}}
\newcommand{\MSTwo}[0]{{M}^S_{2}}
\newcommand{\MSFour}[0]{{M}^S_{3}}
\newcommand{\MSSix}[0]{{M}^S_{4}}
\newcommand{\MSTen}[0]{{M}^S_{5}}
\newcommand{\MSTwelve}[0]{{M}^S_{6}}
\newcommand{\MSFourteen}[0]{{M}^S_{7}}
\newcommand{\MSFifteen}[0]{{M}^S_{8}}
\newcommand{\MSNineteen}[0]{{M}^S_{9}}
\newcommand{\MSTwentyThree}[0]{{M}^S_{10}}
\newcommand{\MSThirtyThree}[0]{{M}^S_{11}}
\newcommand{\MSThirtyFour}[0]{{M}^S_{12}}
\newcommand{\MSThirtySix}[0]{{M}^S_{13}}
\newcommand{\MSThirtySeven}[0]{{M}^S_{14}}

\def\beq{\begin{equation}}
\def\eeq{\end{equation}}
\def\bsp#1\esp{\begin{split}#1\end{split}}

\newcommand{\cM}{{\mathcal{M}}}
\newcommand{\cF}{{\mathcal{F}}}
\newcommand{\ord}{{\mathcal{O}}}

\catcode`\@=11
\font\manfnt=manfnt
\def\Watchout{\@ifnextchar [{\W@tchout}{\W@tchout[1]}}
\def\W@tchout[#1]{{\manfnt\@tempcnta#1\relax%
  \@whilenum\@tempcnta>\z@\do{%
    \char"7F\hskip 0.3em\advance\@tempcnta\m@ne}}}
\let\foo\W@tchout
\def\dubious{\@ifnextchar[{\@dubious}{\@dubious[1]}}

\def\@dubious[#1]{%
  \setbox\@tempboxa\hbox{\@W@tchout#1}
  \@tempdima\wd\@tempboxa
  \list{}{\leftmargin\@tempdima}\item[\hbox to 0pt{\hss\@W@tchout#1}]}
\def\@W@tchout#1{\W@tchout[#1]}
\catcode`\@=12

\title{Soft Expansion of Double-Real-Virtual Corrections to Higgs Production at N$^3$LO}

\author{Charalampos Anastasiou$^a$, Claude Duhr$^{b,c}$\footnote{On leave from the ``Fonds National de la Recherche Scientifique'' (FNRS), Belgium.}\,\,, Falko Dulat$^a$, Elisabetta Furlan$^d$,
Franz Herzog$^{e}$, Bernhard Mistlberger$^a$\\
{}$^a$Institute for Theoretical Physics, ETH Z\"urich,
  8093 Z\"urich, Switzerland
\\
{}$^b$CERN Theory Division, CH-1211, Geneva 23, Switzerland\\
{}$^c$Center for Cosmology, Particle Physics and Phenomenology (CP3),\\
\phantom{{}$^c$}Universit\'e catholique de Louvain,\\
\phantom{{}$^c$}Chemin du Cyclotron 2, 1348 Louvain-La-Neuve, Belgium
\\
{}$^d$Fermilab, Batavia, IL 60510, USA
\\
{}$^e$Nikhef, Science Park 105, NL-1098 XG Amsterdam, The Netherlands
}

\preprint{CERN-PH-TH-2015-092, CP3-15-11, FERMILAB-PUB-15-089-T, NIKHEF/2015-016}

\abstract{
We present methods to compute higher orders in the threshold expansion 
for the one-loop production of a Higgs boson in association with two partons
at hadron colliders. This process contributes to the N$^3$LO Higgs production cross
section beyond the soft-virtual approximation.
We use reverse unitarity to expand the phase-space integrals in the small kinematic
parameters and to reduce the coefficients of the expansion to a small set of master integrals.
We describe two methods for the calculation of the master integrals. The first was
introduced for the calculation of the soft triple-real radiation relevant to N$^3$LO
Higgs production. The second uses
a particular factorization of the three body phase-space measure
and the knowledge of the scaling properties of
the integral itself. Our result is presented as a Laurent expansion
in the dimensional regulator, although some of the master integrals are computed to all orders
in this parameter.

}


\begin{document}


\section{Introduction}
\label{sec:introduction}
The discovery of the Higgs boson~\cite{HiggsDiscovery} opens the path to the exploration of 
the sector responsible for the breaking of the electroweak symmetry 
and makes the Standard Model a fully predictive theory, because 
all couplings are now uniquely fixed. 
Precision measurements of the properties of the Higgs boson 
therefore represent a crucial test of the Standard Model, and 
any deviation, however small, will open a window to new physics
scenarios. 

The main production mechanism of a Higgs boson at hadron colliders
is gluon fusion.
Unfortunately, gluon-fusion suffers from large perturbative instabilities~\cite{nlo,Anastasiou:2002yz,nnlo,ihixs}, 
and the uncertainty on the next-to-next-to leading order (NNLO) prediction 
still remains of the order of $10\%$. In order to fully exploit the potential
of the LHC, it is therefore crucial to improve on the NNLO result by going
to the next order in the perturbative expansion. 

Recently, the 
next-to-next-to-next to leading order (N$^3$LO) 
QCD correction to Higgs boson production has been computed as an expansion around threshold, and it was shown that the remaining QCD scale uncertainty is reduced to a small $2\%$~\cite{Anastasiou:2015ema}. 
Most of the ingredients and subprocesses that go into the computation of ref.~\cite{Anastasiou:2015ema}
have been published separately over the last few years: the three-loop corrections to Higgs
production in gluon fusion, as well as the corrections from the emission of an
additional parton at one or two loops, have been computed in full generality in
refs.~\cite{formfactor,Anastasiou:2013mca,Kilgore:2013gba,Duhr:2014nda,Duhr:2013msa}. In
order to obtain a finite result, appropriate ultra-violet and infrared
counterterms need to be included~\cite{UV,IR,NNLOXsec}. 
The contributions from the emission of two partons at one loop and three
partons at tree-level, however, have not been computed for arbitrary kinematics so far, but they are only known as an expansion around threshold. In particular, in refs.~\cite{triplereal,Anastasiou:2014vaa,Li:2014bfa,Li:2014afw,Anastasiou:2014lda} the first two terms in the threshold expansion of the triple-real and double-real-virtual corrections have been computed by reducing the corresponding phase-space and loop integrals to a small set of master integrals, all of which belong to a special class of integrals that we call \emph{soft integrals}. These soft integrals are not only important when computing the first two terms in the threshold expansion of the cross section, but they also contribute to the full result for the cross section as boundary conditions to the differential equations satisfied by the master integrals in full kinematics. 
Despite their importance, the results for the individual soft integrals that contribute to the soft-virtual and next-to-soft corrections at N$^3$LO of refs.~\cite{Anastasiou:2014vaa,Anastasiou:2014lda} have never been published explicitly\footnote{Some of the soft integrals contributing to the soft-virtual corrections to the cross section at N$^3$LO have been published in ref.~\cite{Li:2014bfa} after the computation of ref.~\cite{Anastasiou:2014vaa} had been completed.}.

The purpose of this paper is threefold: first, we close the aforementioned gap and we present all the technical details that went into the computation of the N$^3$LO cross section in the soft-virtual and next-to-soft approximations of refs.~\cite{Anastasiou:2014vaa,Anastasiou:2014lda}. We discuss in detail methods to perform the threshold expansion of one-loop matrix elements for the production of a heavy colorless state in association with two partons to any desired order, at least in principle. Second, we present techniques to compute the resulting soft master integrals analytically as a Laurent expansion in the dimensional regulator with coefficients that are polynomials in multiple zeta values. Finally, we give explicitly all the soft master integrals contributing to the results of refs.~\cite{Anastasiou:2014vaa,Anastasiou:2014lda}, as well as several new soft master integrals which only contribute at higher orders in the threshold expansion~\cite{Anastasiou:2015ema}.

In a nutshell, our approach to threshold expansion can be described as follows: we start by expanding both the phase-space measure and the interference amplitudes 
into a power series in some small parameter quantifying the deviation from threshold. 
At every order in the expansion, the phase-space integrals are mapped onto (cut) Feynman integrals
by virtue of reverse unitarity~\cite{Anastasiou:2002yz,Anastasiou:2002wq}, which identifies on-shell phase-space constraints with cuts of Feynman propagators. 
The resulting cut integrals are then reduced to a small set of soft master integrals using integration-by-parts (IBP) techniques~\cite{Tkachov:1981wb}.
Similarly, the expansion of the one-loop matrix elements in the integrand is performed using the strategy of regions~\cite{Beneke:1997zp}, which allows to exchange the threshold expansion and the loop integration, provided that contributions from all `regions' in loop momentum space that lead to singularities in the soft limit are included. Similar to the findings of ref.~\cite{Anastasiou:2013mca}, we observe that in the present case the relevant regions correspond to regions where the loop momentum can be either `hard', `soft' or `collinear' to one of the initial-state momenta. While the IBP reduction of the hard and soft regions can be dealt with using standard techniques, the collinear regions require some special attention, because the resulting integrals contain propagators which are non-linear in the Lorentz invariants, and they are hence not amenable to standard techniques.
We find that, due to the analytic structure 
of these propagators, it is always possible to recast the collinear-type integrals into a more standard 
form using partial fractioning, and we observe that the ensuing master 
integrals are always soft integrals, independently of the region. 

In a second part of the paper we discuss techniques to evaluate the soft master integrals as Laurent expansions in the dimensional regulator, and we present two complementary approaches to achieve this. 
The first technique is reminiscent of the method of ref.~\cite{triplereal} and allows one to obtain a multifold Mellin-Barnes representation with poles at integer values for each soft master integral. 
The second method uses a particular factorisation of the three-body phase-space measure, similar to the method presented in ref.~\cite{Zhu:2014fma} for purely real emissions. 
This method allows to arrive at a simple parametric integral representation, whose simplicity can be understood via the scale invariance properties of soft integrals.

The paper is organised as follows. In Section~\ref{sec:Setup} we describe the setup of the 
problem and introduce some notation. In Section~\ref{sec:SoftExpansion} we outline the methods
used for the expansion 
around threshold and the technique of reverse unitarity. We 
discuss the reduction to master integrals and list the complete set of master integrals for each region
in Section~\ref{sec:IBPandSoftMasters}. In Section~\ref{sec:master_computation} we describe the methods used for their evaluation and 
we present analytical results, either as an expansion in the dimensional regulator or 
exactly in terms of generalised hypergeometric functions. We draw our conclusions 
in Section~\ref{sec:conclusions}.


\section{Double-real-virtual corrections to Higgs production}
\label{sec:Setup}
In this paper we consider the production of a Higgs boson ($H$) in association with two massless partons ($k$ and $l$)
via the scattering of two massless partons in the initial state ($i$ and $j$),
\beq
i(q_1)+j(q_2)\to H(q_H)+k(q_3)+l(q_4)\;.
\eeq
Here the partons $i,j,k$ and $l$ can be gluons or massless quarks of $N_F$ different flavours, which are not directly coupled to the Higgs boson. 
We work in the large top-mass limit, and we assume the gluons to couple directly to the Higgs boson via the effective operator
\beq
\mathcal{L}_{\textrm{eff}} = -\frac{1}{4}\,C\,H\,G_{\mu\nu}^a\,G^{\mu\nu}_a\,.
\eeq
At N$^3$LO all of these partonic processes enter the inclusive cross section in the large top-mass limit through 
the interference of the relevant tree and one-loop amplitudes. To be more concrete, let us write the contribution to the cross section as
\beq
\sigma^{\mathrm{RRV}}_{ij\to Hkl}(s,z,\eps)= \frac{\N_{ij;kl}}{2s_{}} C^{(1)}_{ij\to klH}(z,\eps) \,,
\eeq
where $\N_{ij;kl}$ contains all averaging and symmetry factors and $s\equiv s_{12}$ denotes the square of the partonic center of mass energy. The mass of the Higgs boson will be denoted by $m_H$, and for later convenience we introduce the following notations,
\beq
s_{i_1\ldots i_k}\equiv q_{i_1\ldots i_k}^2\,,\qquad q_{i_1\ldots i_k}\equiv\tau_{i_1}q_{i_1}+\ldots+ \tau_{i_k}q_{i_k}\,,
\eeq
where
\beq
\tau_i = 
\left\{ 
\begin{array}{ll} 
+1      &\quad\mbox{if  $i = 1,2$}\,,\\
-1      &\quad\mbox{if  $i > 2$}\,, 
\end{array} 
\right. 
\eeq
and 
\beq
q_H^2=m_H^2,\qquad z=\frac{m_H^2}{s},\qquad \bar z=1-z\; .
\eeq
The dimensionless (coefficient) function $C^{(1)}_{ij\to klH}$ is defined as
\beq\label{eq:PSAA}
 C_{ij\to klH}^{(1)}(z,\eps) = \int \d \Phi_3 \sum_{\text{spins$,$colors}}  2\mathrm{Re} \l( \A_{ij\to H kl}^{(0)}\A_{ij\to H kl}^{(1)*} \r)\,.
\eeq
If we set the scale introduced by dimensional regularisation $\mu$ equal to $m_H$ (a convention which we shall adopt throughout this paper), then the coefficient function only depends on $z$ and the dimensional regulator $\eps$, related to the dimension of space-time $D$ via $\eps=\frac{4-D}{2}$. Here $\A_{ij\to H kl}^{(L)}$ denotes the $L$-loop amplitude for the the process $ij\to H kl$ and the differential three-particle phase-space measure 
is defined as 
\beq\label{eq:PS_measure}
\d\Phi_3= (2\pi)^{D}\delta^{(D)}(q_{1234}-q_H)  \frac{\d^Dq_H}{(2\pi)^{D-1}}\delta_+(q_H^2-m_H^2) \prod_{i=3}^4 \frac{\d^Dq_i}{ (2\pi)^{D-1}}\delta_+(q_i^2)  \; ,
\eeq
where $\delta_+(q^2-m^2)\equiv\delta(q^2-m^2)\,\Theta(q^0)$, with $\Theta$ the Heaviside step function.
In the rest of this paper we present methods to compute the coefficient function $C^{(1)}_{ij\to klH}(z)$ as an expansion around threshold.


\section{Threshold expansion for real emissions at one loop}
\label{sec:SoftExpansion}
Our goal is to obtain the expansion of the coefficient function $C^{(1)}_{ij\to klH}(z,\eps)$ close to $z=1$, where all the emitted final-state partons are soft. In other words, we want to expand eq.~\eqref{eq:PSAA} in the soft momenta $q_3$ and $q_4$. To be more concrete, let us define a set of rescaled momenta $p_i$ as
\beq\label{eq:rescaled_momenta}
q_i = \left\{\begin{array}{rl}
\bar{z}\,p_i\,,&\textrm{ if } i\in\{3,4\}\,,\\
p_i\,,&\textrm{ otherwise}\,.
\end{array}\right.
\eeq
After this rescaling both the phase-space measure and the matrix element in eq.~\eqref{eq:PSAA} depend on the scaling parameter $\bar{z}$. It will be convenient to introduce a notation for how a given quantity scales with $\bar{z}$, e.g., we will write $q_1\sim 1$ and $q_3\sim \bar{z}$ to refer to the scaling behaviour defined by eq.~\eqref{eq:rescaled_momenta}.
Our goal is to expand both the phase-space measure and the amplitudes in this scaling parameter. In the rest of this section we discuss the expansion of each of these objects in turn.

Let us start by discussing the threshold expansion of the phase-space measure. Changing variables to the rescaled momenta~\eqref{eq:rescaled_momenta}, it is obvious that the only quantity in eq.~\eqref{eq:PS_measure} that depends non-trivially on the scaling parameter is the on-shell condition for the Higgs boson,
 \bea
 \delta_+(p_H^2-m_H^2) &=& \delta_+((p_{12}-\bar z p_{34})^2-m_H^2)
                     = \frac{1}{\bar{z}}\,\delta_+(s_{12}-2\,p_{12}\cdot p_{34}+\bar z\,s_{34})\,.
 \eea
 At leading order in the threshold expansion we can simply ignore the term linear in $\bar{z}$ appearing inside the $\delta$ function. If we want to study the subleading terms in the expansion, however, we need to take into account this term, i.e., we need to expand the $\delta$ function around threshold. This can be achieved using \emph{reverse unitarity}~\cite{Anastasiou:2002yz,Anastasiou:2002wq} to interpret phase-space integrals as Feynman integrals with cut-propagators, i.e., where some of the propagators have been replaced by on-shell $\delta$-functions,
  \beq
 \delta_+(q^2)\to \l(\frac{1}{q^2}\r)_c\; .
 \eeq
 The subscript $c$ is a reminder that this propagator is cut.
 Cut propagators can be differentiated just like normal propagators,
 \beq
  \frac{\d}{\d x} \l(\frac{1}{F(x)}\r)_c^n = -n \l(\frac{1}{F(x)}\r)_c^{n+1} \frac{\d F(x)}{\d x}\,,
 \eeq
 but satisfy the extra condition
 \beq
  \l(\frac{1}{q^2}\r)_c^n (q^2)^m=\left\{ 
\begin{array}{lll} 
 \l(\frac{1}{q^2}\r)_c^{n-m}   \,,    &\quad\mbox{if  $n>m$ }\,,\\
0\,,      &\quad\mbox{if  $m\geq n$ }\,,\\ 
\end{array} 
\right. 
 \eeq
 The soft expansion of the phase-space measure therefore reads,
 \beq\label{eq:PS_expand}
 \d \Phi_3= {\bar z}^{3-4\eps} \sum_{n=0}^\infty  \d \Phi_3^{(S,n)} \bar{z}^n\,,
 \eeq
 with
 \beq
 \d \Phi_3^{(S,n)} = 2\pi\, (-s_{34})^n\l( \frac{1}{s_{12}-2p_{12}\cdot p_{34}}\r)_c^{1+n} \prod_{i=3}^4 \frac{\d^Dp_i}{ (2\pi)^{D-1}} \l(\frac{1}{p_i^2}\r)_c\, .
 \eeq
 Note that this procedure necessarily introduces cut propagators raised to higher powers in the subleading terms in the expansion. In Section~\ref{sec:IBPandSoftMasters} we will see that integrals involving these additional powers of cut propagators can always be reduced to the case $n=0$. As a result, all phase-space integrations will be performed against the \emph{soft phase-space measure} defined by
 \beq\label{eq:soft_PS_measure}
\d \Phi_3^{S}\equiv\d \Phi_3^{(S,0)} = 2\pi\, \delta_+( s_{12}-2p_{12}\cdot p_{34}) \prod_{i=3}^4 \frac{\d^Dp_i}{ (2\pi)^{D-1}} \delta_+(p_i^2)\,.
\eeq
 
Let us now turn to the threshold expansion of the integrand. This requires expanding in $\bar{z}$ both the tree-level and one-loop amplitudes appearing inside the matrix element in eq.~\eqref{eq:PSAA}. The expansion of the tree-level amplitude can easily be obtained by introducing the rescaled momenta~\eqref{eq:rescaled_momenta} and expanding the resulting rational function in $\bar{z}$.

The expansion of the one-loop amplitude, however, is more subtle. In order to understand why, let us mention already now that the coefficient function is not meromorphic at $z=1$, but it can nevertheless be written in the form
\beq
\label{Canaliticstructure}
C^{(1)}_{ij\to klH}(z,\eps)=
\bar z^{-1-4\eps} C^{(1,h)}_{ij\to klH}(z,\eps) 
+\bar z^{-1-5\eps} C^{(1,c)}_{ij\to klH}(z,\eps) 
+\bar z^{-1-6\eps} C^{(1,s)}_{ij\to klH}(z,\eps)\,,
\eeq
where the functions $C^{(1,r)}_{ij\to klH}(z)$ for $r\in\{h,c,s\}$ are holomorphic at $z=1$ and therefore admit a Taylor expansion around this point,
\beq
 C^{(1,r)}_{ij\to klH}(z,\eps)=\sum_{k=0}^\infty C^{(1,r,k)}_{ij\to klH}(\eps)\; \bar z^k  \; .
\eeq
A naive expansion of the loop integrand in $\bar{z}$ fails to reproduce this analytic structure, because the expansion of the loop integrand gives rise to a holomorphic function, and the phase-space measure can only account for the term proportional to $\bar{z}^{-4\eps}$ (see eq.~\eqref{eq:PS_expand}). In other words, the threshold expansion does not commute with the loop integration.

Indeed, it is well-known that for some kinematic limits the loop integration does not commute with expansions of the loop integral, and hence a naive 
 Taylor expansion of the loop integrand  is bound to fail.  
 The strategy of regions instead allows one to recover the correct expansion of the integral~\cite{Beneke:1997zp}.
 In this approach one sums up the expansions around all loop momentum regions which can be identified to lead to singularities in the limit $z\to 1$. 
 In ref.~\cite{Anastasiou:2013mca} it was argued that in the present case these different `regions' can be classified as `hard' ($h$), `soft' ($s$) and `collinear' ($c$), depending on whether the loop momentum is hard, soft or collinear to one of the hard partons. The three different terms in eq.~\eqref{Canaliticstructure} are then associated with these three regions. The collinear region is further 
 decomposed into two different collinear regions, `collinear-1' ($c_1$) and `collinear-2' ($c_2$), where the loop momentum is collinear to $p_1$ or $p_2$. 
 Each region is associated with a specific scaling of the (components of the) loop momentum $k$, similar to the rescaling~\eqref{eq:rescaled_momenta} used to define the threshold expansion of the phase-space measure. In order to define these scalings, we parametrise the momentum flowing through a suitable propagator by
 \beq
 \label{ksudakov}
 k^\mu=\alpha\, p_1^\mu+\beta\, p_2^\mu+k_\perp^\mu\,,
 \eeq 
 where
 \beq
 \label{alphabetta}
 \alpha=\frac{k\cdot p_2}{p_1\cdot p_2},\qquad\qquad \beta=\frac{k\cdot p_1}{p_1\cdot p_2}
 \eeq
 and the measure can then be written in this parametrisation as 
 \beq
\d^Dk=\frac{s_{12}}{2}\,\d\alpha\,\d\beta\, \d^{D-2}k_{\perp}\,.
 \eeq
 The different regions are then associated with the following scalings in $\bar{z}$,
 \begin{enumerate}
  \item[($h$):] $k\sim1$, such that  $\d^Dk\sim 1$,
  \item[($c_1$):] $\alpha\sim\bar{z}$, $k_\perp\sim\sqrt{\bar{z}}$, such that $\d^Dk\sim \bar{z}^{\frac{D}{2}}$,
  \item[($c_2$):] $\beta\sim\bar{z}$, $k_\perp\sim\sqrt{\bar{z}}$, such that $\d^Dk\sim \bar{z}^{\frac{D}{2}}$,
  \item[($s$):] $k\sim \bar{z}$, such that $\d^Dk\sim \bar z^{D}$.
 \end{enumerate}
 The overall scaling of a given region in eq.~\eqref{Canaliticstructure} can then be understood by combining the  
 scalings of phase-space measure, eq.~\eqref{eq:PS_expand}, and loop momentum measure in the different regions. 
 
 At this stage we have to address a critical point in this procedure. The scalings we have just defined are not invariant under the shift symmetry of the loop integral, i.e., the rescaling can only be applied once an adequate shift of the loop momentum has been identified. We solve this issue in the following way: first, we note that, up to swaps of the external momenta $p_3, p_4$, all the loop integrations appearing inside the one-loop amplitude can be mapped to one of the following four pentagon topologies,
 \beq\bsp\label{eq:Pentagons}
& \textrm{Pent}_{n_1,\ldots,n_5}(p_1,-p_3,-p_4,p_2)\,,\\
 &\textrm{Pent}_{n_1,\ldots,n_5}(p_1,p_2,-p_4,-p_3)\,, \\
 &\textrm{Pent}_{n_1,\ldots,n_5}(p_{14},p_4,p_{24},-p_3)\,,\\
  &\textrm{Pent}_{n_1,\ldots,n_5}(-p_3,p_1,p_2,-p_4)\,,
  \esp\eeq
  with
 \beq\bsp\label{eq:pentagon_def}
& \textrm{Pent}_{n_1,\ldots,n_5}(q_1,q_2,q_3,q_4) \\
 &= \int \frac{\d^Dk}{i\pi^{D/2}} \frac{1}{[(k-q_1-q_2)^2]^{n_1}\,[(k-q_2)^2]^{n_2}\,[k^2]^{n_3}\,[(k+q_3)^2]^{n_4}\,[(k+q_3+q_4)^2]^{n_5}}\,.
 \esp\eeq
 
Using the loop momentum scalings defined in the previous paragraph we have checked explicitly that the first 
few terms of the $\bar{z}$ expansions, obtained by 
summing up the expansions of non-vanishing hard, collinear and soft regions of various pentagons, boxes, 
triangles and bubbles defined through eq.~\eqref{eq:Pentagons}, 
reproduce the same expansions as one can obtain from an expansion in Feynman parameter space by the 
methods of ref.~\cite{asy}.
Combined with the expansion of the phase-space measure discussed at the beginning of this section, we 
have therefore obtained a machinery to compute the threshold expansion of the coefficient functions 
$C^{(1)}_{ij\to klH}(z,\eps)$.

Let us conclude this section by making a technical comment about the expansion in the collinear regions. 
The procedure outlined above will effectively lead to an expansion in $\sqrt{\bar{z}}$ in the collinear region, 
because the transverse components of the loop momentum scale like $\sqrt{\bar{z}}$. In other words, there 
seems to be a contradiction to the statement that the coefficient function in the collinear region, 
$C^{(1,c)}_{ij\to klH}(z,\eps)$, is holomorphic at $z=1$. In the following we show that this coefficient 
function in the collinear region is indeed holomorphic at $z=1$, and therefore all terms proportional 
to powers $\bar{z}^{n/2}$, with $n$
odd, must vanish. In order to prove this statement, consider an integral of the form,
\beq\label{eq:trans_int}
\int\frac{\d^Dk}{i\pi^{D/2}}\frac{k_\perp^{\mu_1}k_\perp^{\mu_2}\ldots k_\perp^{\mu_n}}{F(k,p_1,p_2)} \,,
\eeq
where $F$ is a Lorentz invariant function which only depends on the loop momentum $k$ through the scalar products $k^2, k\cdot p_1$ and $k\cdot p_2$. 
Using the scaling properties of the collinear regions and the relations of eq.~\eqref{alphabetta}, it can be seen that all integrals which appear in the $\bar z$ expansions of the collinear regions of the four pentagon topologies defined in eq.~\eqref{eq:Pentagons} indeed fall into this category.
As the only source of $\sqrt{\bar{z}}$ in the expansion are the transverse components of the loop momentum $k_{\perp}$, all terms in the expansion proportional to  $\bar{z}^{n/2}$ with $n$ odd must be proportional to an integral of the type~\eqref{eq:trans_int}. A sufficient condition to prove that $C^{(1,r)}_{ij\to klH}(z,\eps)$ is holomorphic at $z=1$ is therefore that all integrals of the type~\eqref{eq:trans_int} vanish for  odd values of $n$. In Appendix~\ref{app:transverse} we prove that this is indeed the case. More precisely, we prove the following result,
\beq\bsp
\label{transverseintegral}
\int\frac{\d^Dk}{i\pi^{D/2}}&\frac{k_\perp^{\mu_1}k_\perp^{\mu_2}\ldots k_\perp^{\mu_n}}{F(k,p_1,p_2)} \\
&\,= 
\left\{ 
\begin{array}{lll} 
\displaystyle
g_\perp^{\mu_1\mu_2\ldots\mu_n} \frac{1}{\prod_{i=1}^{n/2}(D-4+2i)}\int  \frac{\d^Dk}{i\pi^{D/2}}   \frac{(k_\perp^2)^{n/2}}{F(k,p_1,p_2)}    &\quad\mbox{if  $n$ even,}\\
0      &\quad\mbox{if  $n$ odd, }\\ 
\end{array} 
\right. 
\esp\eeq
where 
\beq
g_\perp^{\mu\nu}=g^{\mu\nu}-\frac{p_1^\mu p_2^\nu+p_1^\nu p_2^\mu}{p_1\cdot p_2}
\eeq
is the metric on the space transverse to the vectors $p_1$ and $p_2$, and
\beq
\label{eq:trans_tensor}
g_\perp^{\mu_1\mu_2\ldots\mu_n}=\frac{1}{2^{n/2}(n/2)!}\sum_{\sigma\in S_n} g_\perp^{\mu_{\sigma(1)}\mu_{\sigma(2)}} g_\perp^{\mu_{\sigma(3)}\mu_{\sigma(4)}}\ldots g_\perp^{\mu_{\sigma(n-1)}\mu_{\sigma(n)}}.
\eeq
We note that eqs.~\eqref{transverseintegral} and ~\eqref{ksudakov} can be used efficiently to reduce any one-loop 
integral, which appears in the collinear region, containing arbitrary powers 
of the Lorentz invariants $k\cdot p_3$ and $k\cdot p_4$ in the numerator, to integrals containing only the Lorentz invariants $k\cdot p_1,k\cdot p_2$ and $k^2$ in numerator and denominator. 


\section{IBP reduction and master integrals}
\label{sec:IBPandSoftMasters}
In the previous section we argued that the threshold expansion of the coefficient function $C^{(1)}_{ij\to klH}$ receives contributions from hard, soft and collinear regions, see eq.~\eqref{Canaliticstructure}. The integrals appearing in the expansion are, however, not independent, but can be reduced to a small set of master integrals using integration-by-parts (IBP) identities~\cite{Tkachov:1981wb}. While IBP identities have been introduced for Feynman integrals rather than phase-space integrals, we can use reverse unitarity to interpret phase-space integrals as Feynman integrals with cut-propagators, to which IBP identities are known to apply. Here we go one step further, and we apply IBP identities combined with reverse-unitarity to the individual hard, soft and collinear regions. As a result, we obtain for each region a small set of hard, soft or collinear master integrals in terms of which the coefficients in the threshold expansion can be expressed. Independently of the region, we will see that the master integrals will fall into the class of so-called \emph{soft} integrals, i.e., integrals with respect to the soft phase-space measure~\eqref{eq:soft_PS_measure} of a function that is independently homogeneous under a rescaling of the initial momenta $p_1$ and $p_2$, as well as under a simultaneous rescaling of all the final-state soft momenta. In the remainder of this section we review the IBP reduction in each region, and we present the analytic results for the master integrals in each region. Details about the computation of the master integrals will be given in Section~\ref{sec:master_computation}.

\subsection{The hard region}
We start by discussing the master integrals coming from the hard region. Since the loop momentum is hard, we can immediately expand in the soft final state momenta under the integral sign, and the IBP reduction for the combined loop and phase-space integral follows the same lines as for the purely real soft emission discussed in ref.~\cite{triplereal}.
We find that the whole contribution from the hard region can be expressed in terms of only two master integrals,
\beq\bsp\nonumber
\cM^H_1 &\,=
  \begin{picture}(130,25) (-3,25)
     \SetWidth{1.0}
    \SetColor{Black}
    \SetWidth{2.0}
    \Arc[double,sep=3.5,clock](81.5,11)(26.005,142.028,37.972)
    \Text(125,35)[lb]{\small{\Black{$1$}}}
    \Arc(81.5,43)(26.005,-142.028,-37.972)
    \Line(102,27)(60,27)
    \Line[arrow,arrowpos=0.5,arrowlength=6.667,arrowwidth=2.667,arrowinset=0.2](102,27)(122,37)
    \Line[arrow,arrowpos=0.5,arrowlength=6.667,arrowwidth=2.667,arrowinset=0.2](102,27)(122,17)
    \Line[dash,dashsize=3](82,45)(82,9)
    \Text(-6,35)[lb]{\small{\Black{$1$}}}
    \Text(125,13)[lb]{\small{\Black{$2$}}}
    \Text(-6,13)[lb]{\small{\Black{$2$}}}
    \Arc(40.5,40.9)(23.947,-144.518,-35.482)
    \Arc[clock](40.5,11)(26.005,142.028,37.972)
    \Line[arrow,arrowpos=0.5,arrowlength=6.667,arrowwidth=2.667,arrowinset=0.2](1,17)(22,27)
    \Line[arrow,arrowpos=0.5,arrowlength=6.667,arrowwidth=2.667,arrowinset=0.2](0,38)(22,27)
  \end{picture}= \int \d\Phi_3^S\,\textrm{Bub}(s_{12}) \,,\\
  \\
  \cM^H_2 &\,=
  \begin{picture}(130,25) (-3,12)
    \SetWidth{2.0}
    \SetColor{Black}
    \Line(85,28)(102,-6)
    \Text(-6,21)[lb]{\small{\Black{$1$}}}
    \Text(-6,-1)[lb]{\small{\Black{$2$}}}
    \Arc(33,22.875)(15.541,-140.548,-39.452)
    \Arc[clock](33,1.875)(17.11,139.444,40.556)
    \Line[arrow,arrowpos=0.5,arrowlength=6.667,arrowwidth=2.667,arrowinset=0.2](1,3)(22,13)
    \Line[arrow,arrowpos=0.5,arrowlength=6.667,arrowwidth=2.667,arrowinset=0.2](0,24)(22,13)
    \Line[dash,dashsize=3](79,42)(79,-18)
    \Text(128,32)[lb]{\small{\Black{$1$}}}
    \Text(128,-10)[lb]{\small{\Black{$2$}}}
    \Line[arrow,arrowpos=0.5,arrowlength=6.667,arrowwidth=2.667,arrowinset=0.2](101,34)(125,34)
    \Line[arrow,arrowpos=0.5,arrowlength=6.667,arrowwidth=2.667,arrowinset=0.2](101,-7)(125,-7)
    \Line(46,13)(102,-7)
    \Line(85,28)(102,34)
    \Line[double,sep=3.5](46,13)(85,28)
    \SetWidth{1.0}
    \SetColor{White}
    \Vertex(88,21){4.123}
    \SetWidth{2.0}
    \SetColor{Black}
    \Line(102,34)(70,5)
  \end{picture}= \int \frac{\d\Phi_3^S}{s_{13}\,s_{24}\,s_{34}}\,\textrm{Bub}(s_{12}) \,.\\
  \esp\eeq
    \vskip 1cm
\noindent The double line denotes the Higgs boson, and the dashed line represents the phase-space cut. All other internal uncut lines are scalar propagators.
Moreover, $\textrm{Bub}(p^2)$ denotes the usual one-loop bubble integral,
\beq\label{eq:Bub_def}
\textrm{Bub}(p^2) = \int \frac{\d^Dk}{i\pi^{D/2}}\,\frac{1}{k^2\,(k+p)^2}= \frac{c_\Gamma}{\eps\,(1-2\eps)}\,(-p^2-i0)^{-\eps}\,.
\eeq
 where we defined the usual loop factor
  \beq\label{eq:cGamma}
  c_\Gamma=\frac{\Gamma(1+\eps)\,\Gamma(1-\eps)^2}{\Gamma(1-2\eps)}\,.
  \eeq
The computation of the master integrals will be discussed in detail in Section~\ref{sec:master_computation}. Note that $\textrm{Bub}(s) = \textrm{Bub}(m_H^2) + \ord(\bar{z})$, and so
the hard master integrals can be written as a bubble integral $\textrm{Bub}(m_H^2)$ multiplying one of the two soft master integrals appearing in the double-real soft corrections at NNLO.

\subsection{The soft region}
Next, we discuss the contribution from the soft region, where $k\sim \bar{z}$. It is straightforward to extend the methods of ref.~\cite{triplereal} to the soft region at one loop. Indeed, since the loop momentum scales in the same way as the final-state soft momenta, IBP reduction proceeds in the same way as for purely real soft emissions at tree-level, except that the loop momentum is not constrained to be on shell. At the end of this procedure, we find the following set of soft master integrals,
\beq\bsp\nonumber
\cMSOne &\,=
  \begin{picture}(130,25) (7,25)
    \SetWidth{1.0}
    \SetColor{Black}
    \SetWidth{2.0}
    \Arc[double,sep=3.5,clock](81.5,11)(26.005,142.028,37.972)
    \Text(125,35)[lb]{\small{\Black{$1$}}}
    \Line(22,27)(43,47)
    \Line(43,47)(61,28)
    \Line(62,28)(42,7)
    \Line(43,7)(22,28)
    \Arc(81.5,43)(26.005,-142.028,-37.972)
    \SetWidth{1.0}
    \SetWidth{2.0}
    \Line(22,27)(45,38)
    \Arc[clock](70.643,12.929)(34.37,118.962,24.168)  \Line[arrow,arrowpos=0.5,arrowlength=6.667,arrowwidth=1.7,arrowinset=0.2](21,46.7)(44,46.7)
\Line[arrow,arrowpos=0.5,arrowlength=6.667,arrowwidth=1.7,arrowinset=0.2](21,7)(42.5,7)
\Line[arrow,arrowpos=0.5,arrowlength=6.667,arrowwidth=1.7,arrowinset=0.2](102,27)(122,37)  \Line[arrow,arrowpos=0.5,arrowlength=6.667,arrowwidth=1.7,arrowinset=0.2](102,27)(122,17)
    \Line[dash,dashsize=3](82,57)(82,-3)
    \Text(13,44)[lb]{\small{\Black{$1$}}}
    \Text(125,13)[lb]{\small{\Black{$2$}}}
    \Text(13,3)[lb]{\small{\Black{$2$}}}
  \end{picture}\\\nonumber\\ \nonumber\\
	\nonumber&= \int \d\Phi_3^S\,\frac{\d^Dk}{i \pi^{D/2}}\frac{1}{(-2 k p_1)k^2(k^2-2 k p_3)(2 k p_2-2 p_2 p_3)}\\
  \cMSTwo &\,=
  \begin{picture}(130,65) (7,26)
    \SetWidth{0.5}
    \SetColor{Black}
    \SetWidth{1.0}
    \SetWidth{2.0}
\Line[arrow,arrowpos=0.5,arrowlength=6.667,arrowwidth=1.7,arrowinset=0.2](21,49)(43,49)    \Line[arrow,arrowpos=0.5,arrowlength=6.667,arrowwidth=1.7,arrowinset=0.2](21,8)(43,8)
\Line[arrow,arrowpos=0.5,arrowlength=6.667,arrowwidth=1.7,arrowinset=0.2](102,29)(122,39)    \Line[arrow,arrowpos=0.5,arrowlength=6.667,arrowwidth=1.7,arrowinset=0.2](102,29)(122,19)
    \Line[dash,dashsize=3](82,59)(82,-1)
    \Text(13,46)[lb]{\small{\Black{$1$}}}
    \Text(125,15)[lb]{\small{\Black{$2$}}}
    \Text(13,5)[lb]{\small{\Black{$2$}}}
        \Text(125,37)[lb]{\small{\Black{$1$}}}
    \Line[double,sep=3](64,29)(98.3,29)
    \Line(43,49)(43,8)
    \Line(43,49)(102,29)
    \Line(43,8)(102,29)
    \Line(43,8)(65,29)
    \Line(43,49)(65,29)
  \end{picture}
\\\nonumber\\ \nonumber\\
	\nonumber&= \int \d\Phi_3^S\,\frac{\d^Dk}{i \pi^{D/2}}\frac{1}{(-2 k p_1)(2 k p_2-2 p_2 p_3-2 p_2 p_4)(k^2-2 k p_3)},\\
    \cMSFour &\, =   \begin{picture}(130,55) (7,32)
   \SetWidth{1.0}
    \SetColor{Black}
    \Text(126,48)[lb]{\small{\Black{$1$}}}
    \SetWidth{2.0}   \Line[arrow,arrowpos=0.5,arrowlength=6.667,arrowwidth=1.7,arrowinset=0.2](21,49)(41,35)
\Line[arrow,arrowpos=0.5,arrowlength=6.667,arrowwidth=1.7,arrowinset=0.2](21,18)(41,35)
    \Line[dash,dashsize=3](83,59)(83,-1)
    \Text(13,48)[lb]{\small{\Black{$1$}}}
    \Text(126,12)[lb]{\small{\Black{$2$}}}
    \Text(13,12)[lb]{\small{\Black{$2$}}}
    \Arc[clock](48.5,17.5)(19.039,113.199,13.671)
    \Arc(59.357,39.214)(18.835,-167.071,-66.06)
    \Arc[double,sep=3,clock](71.051,7.102)(41.004,137.127,40.994)    \Line[arrow,arrowpos=0.5,arrowlength=6.667,arrowwidth=1.7,arrowinset=0.2](102,34)(124,49)    \Line[arrow,arrowpos=0.5,arrowlength=6.667,arrowwidth=1.7,arrowinset=0.2](102,34)(124,18)
    \Arc(77,49.875)(29.614,-109.735,-32.416)
    \Arc[clock](91.645,7.16)(28.768,148.946,68.904)
  \end{picture}\\\nonumber\\ \nonumber\\
	\nonumber&= \int \d\Phi_3^S\,\frac{\d^Dk}{i \pi^{D/2}}\frac{1}{(k^2-2 k p_3-2 k p_4+2 p_3 p_4)k^2},
	\esp\eeq
  \beq\bsp
\nonumber
    \cMSSix &\, =   \begin{picture}(130,55) (7,32)
      \SetWidth{2.0}
    \SetColor{Black}
    \Line(43,54)(102,54)
    \Line(102,13)(42,13)
        \Line(72,54)(72,13)
    \SetWidth{1.0}
    \SetColor{White}
    \Vertex(72,34){5.657}
    \SetWidth{2.0}
    \SetColor{Black}
        \Line[double,sep=3](43.5,53)(101.5,13.7)
    \Text(127,51)[lb]{\small{\Black{$1$}}}
    \SetWidth{1.0}
    \SetWidth{2.0}   \Line[arrow,arrowpos=0.5,arrowlength=6.667,arrowwidth=1.7,arrowinset=0.2](21,54)(43,54)   \Line[arrow,arrowpos=0.5,arrowlength=6.667,arrowwidth=1.7,arrowinset=0.2](21,13)(43,13)
    \Line[dash,dashsize=3](88,59)(88,-1)
    \Text(13,50)[lb]{\small{\Black{$1$}}}
    \Text(127,10)[lb]{\small{\Black{$2$}}}
    \Text(13,10)[lb]{\small{\Black{$2$}}}
    \Line(102,54)(102,13)   \Line[arrow,arrowpos=0.5,arrowlength=6.667,arrowwidth=1.7,arrowinset=0.2](102,54)(124,54)   \Line[arrow,arrowpos=0.5,arrowlength=6.667,arrowwidth=1.7,arrowinset=0.2](102,13)(124,13)
    \Line(43,54)(43,13)
  \end{picture}\\\nonumber\\ \nonumber\\
	\nonumber&= \int \d\Phi_3^S\,\frac{\d^Dk}{i \pi^{D/2}}\frac{1}{(-2 p_1 p_4)(-2 k p_2)(k^2-2 k p_3)(k^2-2 k p_3-2 k p_4+2 p_3 p_4)k^2}\,,\\
  \cMSTen &\, =
  \begin{picture}(130,55) (7,26)
    \SetWidth{1.0}
    \SetColor{Black}
    \SetWidth{2.0}
   \Text(125,35)[lb]{\small{\Black{$1$}}}
    \Line(22,27)(43,47)
    \Line(43,47)(61,28)
    \Line(62,28)(42,7)
    \Line(43,7)(22,28)    \SetWidth{1.0}
    \SetWidth{2.0}
    \Line(23,28)(45,40)
    \Arc[clock](70.643,14)(34.37,118.962,24.168)    \Line[arrow,arrowpos=0.5,arrowlength=6.667,arrowwidth=1.7,arrowinset=0.2](21,46.7)(44,46.7)
\Line[arrow,arrowpos=0.5,arrowlength=6.667,arrowwidth=1.7,arrowinset=0.2](21,7)(42.5,7)
\Line[arrow,arrowpos=0.5,arrowlength=6.667,arrowwidth=1.7,arrowinset=0.2](102,27)(122,37)  \Line[arrow,arrowpos=0.5,arrowlength=6.667,arrowwidth=1.7,arrowinset=0.2](102,27)(122,17)
    \Line[dash,dashsize=3](82,59)(82,-1)
    \Text(13,44)[lb]{\small{\Black{$1$}}}
    \Text(125,13)[lb]{\small{\Black{$2$}}}
    \Text(13,3)[lb]{\small{\Black{$2$}}}
    \Line[double,sep=3](60,27.2)(102,27.2)
    \Line(22,28)(45,17)
    \Arc(69.783,47.5)(38.552,-114.166,-33.314)
  \end{picture}\\\nonumber\\ \nonumber\\
	\nonumber&=\int \d\Phi_3^S\,\frac{\d^Dk}{i \pi^{D/2}}\frac{1}{(k^2+2 k p_3)(k^2-2 k p_4)(2 k p_2-2 p_2 p_4)(-2 p_1 p_3-2 k p_1)}\,,\\
  \cMSTwelve&\,= \begin{picture}(130,55) (7,26)
      \SetWidth{2.0}
    \Line(43,46.7)(102,46.7)
    \Line(22,27.5)(102,8)
        \SetWidth{1.0}
    \SetColor{White}
    \SetWidth{2.0}
    \SetColor{Black}
    \SetWidth{1.0}
    \SetColor{White}
    \Vertex(55,20){5}
    \SetWidth{2.0}
    \SetColor{Black}
    \SetWidth{2.0}
    \SetColor{Black}
       \Text(127,44)[lb]{\small{\Black{$1$}}}
    \Line(22,27)(43,47)
    \Line(43,47)(61,28)
    \Line(62,28)(42,7)
    \Line(43,7)(22,28)
    \SetWidth{1.0}
    \SetWidth{2.0}  \Line[arrow,arrowpos=0.5,arrowlength=6.667,arrowwidth=1.7,arrowinset=0.2](21,46.7)(44,46.7)
\Line[arrow,arrowpos=0.5,arrowlength=6.667,arrowwidth=1.7,arrowinset=0.2](21,7)(42.5,7)
    \Line[dash,dashsize=3](82,59)(82,-1)
    \Text(127,4)[lb]{\small{\Black{$2$}}}
    \Line[double,sep=3](60,27.2)(102,45.6)
    \Line(102,46.7)(102,8)
\Line[arrow,arrowpos=0.5,arrowlength=6.667,arrowwidth=1.7,arrowinset=0.2](101.5,46.7)(124,46.7)   \Line[arrow,arrowpos=0.5,arrowlength=6.667,arrowwidth=1.7,arrowinset=0.2](101.5,8)(124,8)
    \Text(13,44)[lb]{\small{\Black{$2$}}}
    \Text(13,3)[lb]{\small{\Black{$1$}}}
  \end{picture}\\\nonumber\\ \nonumber\\
	\nonumber&=\int \d\Phi_3^S\frac{\d^Dk}{i \pi^{D/2}}\frac{1}{(-2 p_2 p_3)k^2(2 k p_2-2 p_2 p_4)(-2 p_1 p_3-2 k p_1)(k^2+2 k p_3)}\,,\\
    \nonumber
    \cMSFourteen&\,= \begin{picture}(130,55) (7,28)
       \SetWidth{2.0}
    \SetColor{Black}
        \Line(67,13)(102,54)
            \SetColor{White}
                \Vertex(92,42){5.5}
   \SetWidth{1.0}
    \SetColor{Black}
    \Text(127,51)[lb]{\small{\Black{$1$}}}
    \SetWidth{2.0}    \Line[arrow,arrowpos=0.5,arrowlength=6.667,arrowwidth=1.7,arrowinset=0.2](21,49)(41,35)
\Line[arrow,arrowpos=0.5,arrowlength=6.667,arrowwidth=1.7,arrowinset=0.2](21,18)(41,35)
    \Line[dash,dashsize=3](82,59)(82,-1)
    \Text(13,48)[lb]{\small{\Black{$1$}}}
    \Text(127,10)[lb]{\small{\Black{$2$}}}
    \Text(13,14)[lb]{\small{\Black{$2$}}}
    \Line(102,54)(102,13)    \Line[arrow,arrowpos=0.5,arrowlength=6.667,arrowwidth=1.7,arrowinset=0.2](101,54)(124,54)
\Line[arrow,arrowpos=0.5,arrowlength=6.667,arrowwidth=1.7,arrowinset=0.2](101,13)(124,13)
    \Line(67,13)(102,13)
    \Arc[clock](46.196,14.777)(20.88,104.409,-4.882)
    \Arc(61.274,32.597)(20.416,173.24,286.287)
    \Arc[double,sep=3,clock](70,7)(41,137,38.5)
   \end{picture}\\\nonumber\\ \nonumber\\
	\nonumber&=\int \d\Phi_3^S \frac{\d^Dk}{i \pi^{D/2}}\frac{1}{(-2 p_1 p_3)(-2 p_2 p_4)(k^2-2 k p_4)(k^2+2 k p_3)}\,,\\
    \cMSFifteen&\,= \begin{picture}(130,55) (7,35)
    \SetWidth{0.5}
    \SetColor{Black}
    \SetWidth{1.0}
    \SetWidth{2.0}
\Line[arrow,arrowpos=0.5,arrowlength=6.667,arrowwidth=1.7,arrowinset=0.2](21,53)(43,53)    \Line[arrow,arrowpos=0.5,arrowlength=6.667,arrowwidth=1.7,arrowinset=0.2](21,12)(43,12)
\Line[arrow,arrowpos=0.5,arrowlength=6.667,arrowwidth=1.7,arrowinset=0.2](101,33)(122,43)    \Line[arrow,arrowpos=0.5,arrowlength=6.667,arrowwidth=1.7,arrowinset=0.2](100,32)(122,23)
    \Line[dash,dashsize=3](72,59)(72,-1)
    \Line[double,sep=3](59,13)(99.5,33.5)
    \Line(43,53)(43,12)
    \Line(43,53)(102,33)
    \Line(43,12)(60,12)
    \Line(43,12)(84,39)
        \Text(14,50)[lb]{\small{\Black{$1$}}}
    \Text(126,18)[lb]{\small{\Black{$2$}}}
    \Text(14,9)[lb]{\small{\Black{$2$}}}
        \Text(126,40)[lb]{\small{\Black{$1$}}}
    \SetWidth{1.0}
    \SetColor{White}
    \Vertex(56,21){3.606}
    \SetWidth{2.0}
    \SetColor{Black}
    \Line(43,53)(60,12)
  \end{picture}\\\nonumber\\ \nonumber\\
	\nonumber&=\int \d\Phi_3^S\frac{\d^Dk}{i \pi^{D/2}}\frac{1}{k^2(2 k p_2-2 p_2 p_4)(-2 p_1 p_3-2 k p_1)(2 p_3 p_4)}\,,
	\esp\eeq
  \beq\bsp
    \cMSNineteen&\,= \begin{picture}(130,55) (7,28)
    \SetWidth{2.0}
    \SetColor{Black}
    \Line(30.5,19.8)(81,36)
    \SetWidth{1.0}
    \SetColor{White}
    \Vertex(48.5,26){4.472}
    \SetColor{Black}
    \SetWidth{2.0}    \Line[arrow,arrowpos=0.5,arrowlength=6.667,arrowwidth=1.7,arrowinset=0.2](21,49)(43,49)
\Line[arrow,arrowpos=0.5,arrowlength=6.667,arrowwidth=1.7,arrowinset=0.2](21,8)(43.5,8)    \Line[arrow,arrowpos=0.5,arrowlength=6.667,arrowwidth=1.7,arrowinset=0.2](102,30)(122,39)
\Line[arrow,arrowpos=0.5,arrowlength=6.667,arrowwidth=1.7,arrowinset=0.2](102,28)(122,19)
    \Line[dash,dashsize=3](67,59)(67,-1)
    \Text(13,46)[lb]{\small{\Black{$1$}}}
    \Text(125,15)[lb]{\small{\Black{$2$}}}
    \Text(13,5)[lb]{\small{\Black{$2$}}}
        \Text(125,37)[lb]{\small{\Black{$1$}}}
    \Line[double,sep=3](54,20.5)(102,28.8)
    \Line(31,20)(43.5,32.5)
    \Line(43,32)(55,20)
    \Line(55,20)(43,8)
    \Line(43,8)(31,20)
    \Line(43,49)(43,32)
    \Line(43,49)(102,29)
  \end{picture}\\\nonumber\\ \nonumber\\
	\nonumber&=\int \d\Phi_3^S\frac{\d^Dk}{i \pi^{D/2}}\frac{1}{(2 p_3 p_4)k^2(-2 p_1 p_3)(2 k p_2-2 p_2 p_4)(-2 p_1 p_3-2 k p_1)(k^2-2 k p_4)}\,,\\
  \cMSTwentyThree&\,= \begin{picture}(130,55) (7,26)
      \SetWidth{2.0}
      \Line(22,27.5)(40,27.5)
    \Line(40,27.5)(102,46.7)
    \Line(40,27.5)(102,8)
        \SetWidth{1.0}
    \SetColor{White}
    \Vertex(56,33){5}
    \SetWidth{2.0}
    \SetColor{Black}
    \SetWidth{1.0}
    \SetColor{White}
    \Vertex(57,23){5}
    \SetWidth{2.0}
    \SetColor{Black}
    \SetWidth{2.0}
    \SetColor{Black}
       \Text(127,44)[lb]{\small{\Black{$1$}}}
    \Line(22,27)(43,47)
    \Line(43,47)(61,28)
    \Line(62,28)(42,7)
    \Line(43,7)(22,28)
    \SetWidth{1.0}
    \SetWidth{2.0}  \Line[arrow,arrowpos=0.5,arrowlength=6.667,arrowwidth=1.7,arrowinset=0.2](21,46.7)(44,46.7)
\Line[arrow,arrowpos=0.5,arrowlength=6.667,arrowwidth=1.7,arrowinset=0.2](21,7)(42.5,7)
    \Line[dash,dashsize=3](82,59)(82,-1)
    \Text(127,4)[lb]{\small{\Black{$2$}}}
    \Line[double,sep=3](60,27.2)(102,27.2)
    \Line(102,46.7)(102,8)
\Line[arrow,arrowpos=0.5,arrowlength=6.667,arrowwidth=1.7,arrowinset=0.2](101.5,46.7)(124,46.7)   \Line[arrow,arrowpos=0.5,arrowlength=6.667,arrowwidth=1.7,arrowinset=0.2](101.5,8)(124,8)
    \Text(13,44)[lb]{\small{\Black{$1$}}}
    \Text(13,3)[lb]{\small{\Black{$2$}}}
    \end{picture}\\\nonumber\\ \nonumber\\
	\nonumber&=\int \d\Phi_3^S\frac{\d^Dk}{i \pi^{D/2}}\frac{1}{(2 k p_2-2 p_2 p_4)(-2 p_1 p_3-2 k p_1)(k^2-2 k p_4)(k^2+2 k p_3)}\,\\
	\nonumber &\times \frac{1}{(2 p_3 p_4)(-2 p_1 p_3)(-2 p_2 p_4)}\\
     \nonumber
    \cMSThirtyThree &\, =   \begin{picture}(130,55) (7,32)
        \SetWidth{2.0}
    \SetColor{Black}
    \Line(43,54)(102,54)
    \Line(102,13)(42,13)
    \Line(73,54)(73,13)
    \Text(127,51)[lb]{\small{\Black{$2$}}}
    \SetWidth{1.0}
    \SetWidth{2.0}
    \Line[arrow,arrowpos=0.5,arrowlength=6.667,arrowwidth=2.667,arrowinset=0.2](21,54)(43,54)
    \Line[arrow,arrowpos=0.5,arrowlength=6.667,arrowwidth=2.667,arrowinset=0.2](21,13)(43,13)
    \Text(13,51)[lb]{\small{\Black{$1$}}}
    \Text(127,11)[lb]{\small{\Black{$1$}}}
    \Text(13,11)[lb]{\small{\Black{$2$}}}
    \Line(102,54)(102,13)
    \Line[arrow,arrowpos=0.5,arrowlength=6.667,arrowwidth=2.667,arrowinset=0.2](102,54)(124,54)
    \Line[arrow,arrowpos=0.5,arrowlength=6.667,arrowwidth=2.667,arrowinset=0.2](102,13)(124,13)
    \Line(43,54)(43,13)
    \SetWidth{1.0}
    \SetColor{White}
    \Vertex(72,40){5.657}
    \Vertex(72,40){5.657}
    \SetWidth{2.0}
    \SetColor{Black}
    \Arc[double,sep=3.5](73,78.667)(38.667,-136.397,-43.603)
    \Line[dash,dashsize=3](88,59)(88,-1)
     \end{picture}\\\nonumber\\ \nonumber\\
	\nonumber&= \int \d\Phi_3^S\frac{\d^Dk}{i \pi^{D/2}}\frac{1}{(k^2+2 k p_3)k^2(2 k p_2-2 p_2 p_4)(k^2-2 k p_4)(-2 p_1 p_4)}\,,\\
     \nonumber
    \cMSThirtyFour &\, =   \begin{picture}(130,55) (7,32)
      \SetWidth{2.0}
    \SetColor{Black}
    \Line(43,54)(102,54)
    \Line(102,13)(42,13)
        \Line(72,54)(72,13)
    \SetWidth{1.0}
    \SetColor{White}
    \Vertex(72,34){5.657}
    \SetWidth{2.0}
    \SetColor{Black}
        \Line[double,sep=3](43.5,53)(101.5,13.7)
    \Text(127,51)[lb]{\small{\Black{$2$}}}
    \SetWidth{1.0}
    \SetWidth{2.0}   \Line[arrow,arrowpos=0.5,arrowlength=6.667,arrowwidth=1.7,arrowinset=0.2](21,54)(43,54)   \Line[arrow,arrowpos=0.5,arrowlength=6.667,arrowwidth=1.7,arrowinset=0.2](21,13)(43,13)
    \Line[dash,dashsize=3](88,59)(88,-1)
    \Text(14,51)[lb]{\small{\Black{$1$}}}
    \Text(127,11)[lb]{\small{\Black{$1$}}}
    \Text(14,11)[lb]{\small{\Black{$2$}}}
    \Line(102,54)(102,13)   \Line[arrow,arrowpos=0.5,arrowlength=6.667,arrowwidth=1.7,arrowinset=0.2](102,54)(124,54)   \Line[arrow,arrowpos=0.5,arrowlength=6.667,arrowwidth=1.7,arrowinset=0.2](102,13)(124,13)
    \Line(43,54)(43,13)
     \end{picture}\\\nonumber\\ \nonumber\\
	\nonumber&= \int \d\Phi_3^S\frac{\d^Dk}{i \pi^{D/2}}\frac{1}{(k^2+2 k p_3)k^2(2 k p_2-2 p_2 p_4)(k^2-2 k p_4)(-2 p_2 p_3)}\,,\\
            \nonumber
    \cMSThirtySix &\, =   \begin{picture}(130,55) (7,32)
    \SetWidth{2.0}
    \SetColor{Black}
    \Line(43,54)(102,54)
    \Line(102,13)(42,13)
    \Line(73,54)(73,13)
    \Text(127,51)[lb]{\small{\Black{$2$}}}
    \SetWidth{1.0}
    \SetWidth{2.0}
    \Line[arrow,arrowpos=0.5,arrowlength=6.667,arrowwidth=2.667,arrowinset=0.2](21,54)(43,54)
    \Line[arrow,arrowpos=0.5,arrowlength=6.667,arrowwidth=2.667,arrowinset=0.2](21,13)(43,13)
    \Text(13,51)[lb]{\small{\Black{$1$}}}
    \Text(127,11)[lb]{\small{\Black{$1$}}}
    \Text(13,11)[lb]{\small{\Black{$2$}}}
    \Line(102,54)(102,13)
    \Line[arrow,arrowpos=0.5,arrowlength=6.667,arrowwidth=2.667,arrowinset=0.2](102,54)(124,54)
    \Line[arrow,arrowpos=0.5,arrowlength=6.667,arrowwidth=2.667,arrowinset=0.2](102,13)(124,13)
    \Line(43,54)(43,13)
    \SetWidth{1.0}
    \SetColor{White}
    \Vertex(73,34){5.657}
    \SetWidth{2.0}
    \SetColor{Black}
    \Line[dash,dashsize=3](88,59)(88,-1)
    \Line[double,sep=3.5](43,34)(102,34)
     \end{picture}\\\nonumber\\ \nonumber\\
	\nonumber& = \int \d\Phi_3^S\frac{\d^Dk}{i \pi^{D/2}}\frac{1}{(2 k p_2-2 p_2 p_4)(k^2-2 k p_4)(k^2+2 k p_3)(-2 p_2 p_3)k^2(-2 p_1 p_4)(-2 p_1 p_3-2 k p_1)}\,,
	\esp\eeq
  \beq\bsp
            \nonumber
    \cMSThirtySeven &\, =   \begin{picture}(130,55) (7,32)
   \SetWidth{1.0}
    \SetColor{Black}
 \SetColor{White}
    \Vertex(59,35){3.606}
    \SetWidth{2.0}
    \SetColor{Black}
    \Line(43,12)(61,12)
    \Line(43,53)(61,53)
    \Line(43,33)(86,40)
    \Line[dash,dashsize=3](72,59)(72,-1)
    \Line[double,sep=3](60,12)(102,33)
    \Line(43,53)(43,12)
    \Line(60,53)(102,33)
    \Text(126,43)[lb]{\small{\Black{$1$}}}
    \Line[arrow,arrowpos=0.5,arrowlength=6.667,arrowwidth=2.667,arrowinset=0.2](21,53)(43,53)
    \Line[arrow,arrowpos=0.5,arrowlength=6.667,arrowwidth=2.667,arrowinset=0.2](21,12)(43,12)
    \Line[arrow,arrowpos=0.5,arrowlength=6.667,arrowwidth=2.667,arrowinset=0.2](102,33)(122,43)
    \Line[arrow,arrowpos=0.5,arrowlength=6.667,arrowwidth=2.667,arrowinset=0.2](102,33)(122,23)
    \SetColor{White}
    \Vertex(60,36){5.657}
    \SetWidth{2.0}
    \SetColor{Black}
    \Line(60,53)(60,12)
    \Text(13,51)[lb]{\small{\Black{$2$}}}
    \Text(126,18)[lb]{\small{\Black{$2$}}}
    \Text(13,9)[lb]{\small{\Black{$1$}}}
  \end{picture}\\\nonumber\\ \nonumber\\
	\nonumber&= \int \d\Phi_3^S\frac{\d^Dk}{i \pi^{D/2}}\frac{1}{k^2(-2 k p_1)(2 k p_2-2 p_2 p_4)(k^2-2 k p_4)(2 p_3 p_4)(2 k p_2-2 p_2 p_3-2 p_2 p_4)}\,.\esp\eeq
  \vskip 1cm
\noindent 
The number of propagators of the soft master integrals depending on the loop momenta ranges from two to five.
Since we are working in the soft region, the loop integrals are not the complete momentum space integrals, but they correspond to the eikonal approximation.
Note that the eikonal approximation of the loop integrals depends on the specific orientation with which it enters the soft phase-space integrals, or equivalently, which invariants become soft. We have explicitly evaluated all loop integrals in the eikonal approximation by determining the scaling of the Feynman parameters in the limit using the package {\tt asy.m}~\cite{asy}. We find that in all cases, except for one pentagon integral, the remaining parametric integrations can be perfomed in closed form to all orders in the dimensional regulator $\eps$. For the remaining pentagon we are able to obtain Mellin-Barnes integral representation valid to all orders in $\eps$. In the following we give a brief summary of the results for the soft virtual integrals. The bubble type integrals in $\mathcal{M}^S_3$ and $\mathcal{M}^S_7$ can just be computed as in eq.~\eqref{eq:Bub_def}, and will not be discussed any further.  Note that we present the results in the Euclidean region where all invariants are negative, and the analytic continuation to the scattering region is  given by
\beq
-m_H^2 - i 0  \to e^{-i\pi}\,m_H^2\,,\qquad -s_{12} - i0 \to e^{-i\pi}\,s_{12}\,,\qquad -s_{34} -i0\to e^{-i\pi}\,s_{34}\,.
\eeq
Note also that the loop integrals appearing inside the master integrals have the correct homogeneity properties to turn the master integrals into soft integrals. The computation of the master integrals will be discussed in Section~\ref{sec:master_computation}.

\paragraph{Soft Triangle Integrals.}
The soft master integrals $\mathcal{M}^S_2$ and $\mathcal{M}^S_8$ contain a virtual integral with three propagators.

\beq\bsp\nonumber
\scalebox{0.4}{
  \begin{picture}(120,30) (200,25)
    \SetWidth{0.5}
    \SetColor{Black}
    \Text(120,112)[lb]{\Huge{\Black{$-p_{1234}$}}}
    \SetWidth{4.0}
    \Line(80,-16)(240,-16)
    \Line(240,-16)(160,80)
    \Line(160,80)(80,-16)
    \SetWidth{8.0}
    \Line[arrowpos=0.5,arrowlength=7.5,arrowwidth=8,arrowinset=0.2](48,-32)(80,-16)
    \Line[arrowpos=0.5,arrowlength=7.5,arrowwidth=8,arrowinset=0.2](272,-32)(240,-16)
    \Line[arrowpos=0.5,arrowlength=7.5,arrowwidth=8,arrowinset=0.2](160,96)(160,80)
    \Text(20,-55)[lb]{\Huge{\Black{$p_{13}$}}}
    \Text(270,-55)[lb]{\Huge{\Black{$p_{24}$}}}
  \end{picture}}
&= \int \frac{d^D k}{i \pi^{D/2}} \frac{1}{k^2(-2kp_1-2p_1p_3)(2kp_2-2p_2p_4)} \\ \nonumber\\
&= -\frac{\Gamma(1-\eps)\,\Gamma(1+\eps)^2}{\eps^2}\,(-s_{13})^{-\eps}\,(-s_{24})^{-\eps}\,(-s_{12})^{-1+\eps}\,.
\esp\eeq
The integral corresponds to a completely off-shell triangle where two of the virtualities of the external momenta ($s_{24}=p_{24}^2$ and $s_{13}=p_{13}^2$) scale as $\bar{z}$.

\paragraph{Soft Box Integrals.}
We observe that there are four configurations of loop integrals with four propagators. Two of them arise from box master integrals with one off-shell leg and two from configurations with two off-shell legs.
\begin{enumerate}
\item The loop integral appearing in the definition of $\mathcal{M}^S_1$ is given by
\beq\bsp
\scalebox{0.4}{
   \begin{picture}(240,108) (19,105)
    \SetWidth{4.0}
    \SetColor{Black}
    \Line(192,32)(192,160)
    \Line(48,160)(48,32)
    \Line(48,32)(192,32)
    \Line(192,160)(48,160)
    \SetWidth{4.0}
    \Line[arrowpos=0.5,arrowlength=7.5,arrowwidth=4,arrowinset=0.2](208,176)(192,160)
    \Line[arrowpos=0.5,arrowlength=7.5,arrowwidth=4,arrowinset=0.2](32,176)(48,160)
    \Line[arrowpos=0.5,arrowlength=7.5,arrowwidth=4,arrowinset=0.2](32,16)(48,32)
    \SetWidth{8.0}
    \Line[arrowpos=0.5,arrowlength=5,arrowwidth=8,arrowinset=0.2](208,16)(192,32)
    \Text(15,185)[lb]{\Huge{\Black{$p_3$}}}
    \Text(205,185)[lb]{\Huge{\Black{$p_1$}}}
    \Text(170,-5)[lb]{\Huge{\Black{$-p_{123}$}}}
    \Text(15,-5)[lb]{\Huge{\Black{$p_2$}}}
  \end{picture}}&= \int\frac{d^Dk}{i\pi^{D/2}} \frac{1}{k^2(k-p_3)^2(-2k p_1)(-2p_2 p_3+2kp_2)}\\
  &\nonumber\\
&=\frac{2 (-s_{12})^{\epsilon } (-s_{13})^{-\epsilon -1} (-s_{23})^{-\epsilon -1} \Gamma (1-\epsilon )^3 \Gamma (\epsilon +1)^2}{\epsilon ^2 \Gamma (1-2 \epsilon )}.
\esp\eeq
\item The soft master integrals  $\mathcal{M}^S_{4}$, $\mathcal{M}^S_{11}$ and $\mathcal{M}^S_{12}$ contain the loop integral
\beq\bsp
\scalebox{0.4}{
   \begin{picture}(240,108) (19,105)
    \SetWidth{4.0}
    \SetColor{Black}
    \Line(192,32)(192,160)
    \Line(48,160)(48,32)
    \Line(48,32)(192,32)
    \Line(192,160)(48,160)
    \SetWidth{4.0}
    \Line[arrowpos=0.5,arrowlength=7.5,arrowwidth=4,arrowinset=0.2](208,176)(192,160)
    \Line[arrowpos=0.5,arrowlength=7.5,arrowwidth=4,arrowinset=0.2](32,176)(48,160)
    \Line[arrowpos=0.5,arrowlength=7.5,arrowwidth=4,arrowinset=0.2](32,16)(48,32)
    \SetWidth{8.0}
    \Line[arrowpos=0.5,arrowlength=5,arrowwidth=8,arrowinset=0.2](208,16)(192,32)
    \Text(15,185)[lb]{\Huge{\Black{$p_3$}}}
    \Text(205,185)[lb]{\Huge{\Black{$p_4$}}}
    \Text(170,-5)[lb]{\Huge{\Black{$-p_{234}$}}}
    \Text(15,-5)[lb]{\Huge{\Black{$p_2$}}}
  \end{picture}}
  \nonumber &= \int\frac{d^Dk}{i\pi^{D/2}} \frac{1}{k^2(k-p_3)^2(k-p_3-p_4)^2(-2k p_2)}\\
   &\nonumber\\
&=-\frac{2 (-s_{34})^{-\epsilon -1} \Gamma (1-\epsilon )^2 \Gamma (\epsilon +1) }{\epsilon
   ^2 (s_{23}+s_{24}) \Gamma (1-2 \epsilon )} \,   _2F_1\left(1,1;1-\epsilon ;\frac{s_{24}}{s_{23}+s_{24}}\right),
\esp\eeq
where $_2F_1$ denotes Gauss' hypergeometric function,
\beq
{}_2F_1(a,b;c;z) = \sum_{n=0}^\infty\frac{(a)_n(b)_n}{(c)_n}\,\frac{z^n}{n!}\,.
\eeq
\item In $\mathcal{M}^S_5$ and $\mathcal{M}^S_{10}$ we find a limit of a two-mass-easy box integral.
\beq\bsp
\scalebox{0.4}{
   \begin{picture}(240,108) (19,105)
    \SetWidth{4.0}
    \SetColor{Black}
    \Line(192,32)(192,160)
    \Line(48,160)(48,32)
    \Line(48,32)(192,32)
    \Line(192,160)(48,160)
    \SetWidth{4.0}
    \Line[arrowpos=0.5,arrowlength=7.5,arrowwidth=4,arrowinset=0.2](208,176)(192,160)
    \Line[arrowpos=0.5,arrowlength=7.5,arrowwidth=4,arrowinset=0.2](32,16)(48,32)
    \SetWidth{8.0}
    \Line[arrowpos=0.5,arrowlength=7.5,arrowwidth=8,arrowinset=0.2](32,176)(48,160)
    \Line[arrowpos=0.5,arrowlength=7.5,arrowwidth=8,arrowinset=0.2](208,16)(192,32)
    \Text(15,185)[lb]{\Huge{\Black{$p_{34}$}}}
    \Text(205,185)[lb]{\Huge{\Black{$p_1$}}}
    \Text(170,-5)[lb]{\Huge{\Black{$-p_{1234}$}}}
    \Text(15,-5)[lb]{\Huge{\Black{$p_2$}}}
  \end{picture}}
  \nonumber &= \int\frac{d^Dk}{i\pi^{D/2}} \frac{1}{(k+p_3)^2(k-p_4)^2(2 k p_2-2p_2 p_4)(-2p_1 p_3-2 k p_1)}\\
   &\nonumber\\
=-(-s_{34})^{-\epsilon } &\frac{2 \Gamma (1-\epsilon )^2 \Gamma (\epsilon +1) }{(\epsilon +1)^2 (s_{13}+s_{14}) (s_{23}+s_{24}) \Gamma (1-2 \epsilon )}\\
\nonumber \times \, _2F_1&\left(1,1;\epsilon +2;1-\frac{s_{12}s_{34}}{(s_{13}+s_{14}) (s_{23}+s_{24})}\right).
\esp\eeq
\item The four propagator loop integral appearing in master integral $\mathcal{M}^S_{6}$ is a limit of a two-mass-hard box integral.
\beq\bsp
\scalebox{0.4}{
   \begin{picture}(240,108) (19,105)
    \SetWidth{4.0}
    \SetColor{Black}
    \Line(192,32)(192,160)
    \Line(48,160)(48,32)
    \Line(48,32)(192,32)
    \Line(192,160)(48,160)
    \SetWidth{4.0}
    \Line[arrowpos=0.5,arrowlength=7.5,arrowwidth=4,arrowinset=0.2](32,176)(48,160)
    \Line[arrowpos=0.5,arrowlength=7.5,arrowwidth=4,arrowinset=0.2](32,16)(48,32)
    \SetWidth{8.0}
    \Line[arrowpos=0.5,arrowlength=7.5,arrowwidth=8,arrowinset=0.2](208,176)(192,160)
    \Line[arrowpos=0.5,arrowlength=7.5,arrowwidth=8,arrowinset=0.2](208,16)(192,32)
    \Text(15,185)[lb]{\Huge{\Black{$p_3$}}}
    \Text(205,185)[lb]{\Huge{\Black{$p_{24}$}}}
    \Text(170,-5)[lb]{\Huge{\Black{$-p_{1234}$}}}
    \Text(15,-5)[lb]{\Huge{\Black{$p_1$}}}
  \end{picture}} \nonumber &= \int\frac{d^Dk}{i\pi^{D/2}} \frac{1}{k^2(k+p_3)^2(-2p_1 p_3 - 2 k p_1)(2k p_2-2 p_2 p_4)}\\
   &\nonumber\\
=(-s_{12})^{\epsilon } (-s_{13})^{-\epsilon -1} (-s_{24})^{-\epsilon -1} &\frac{ \Gamma
   (1-\epsilon ) \Gamma (\epsilon +1)^2 }{\epsilon ^2}\, _2F_1\left(1,\epsilon +1;1-\epsilon
   ;-\frac{s_{23}}{s_{24}}\right).
\esp\eeq
The loop integral in $\mathcal{M}^S_{9}$ is related to the integral above via the permutation of external momenta $(p_1\rightarrow p_2,p_2\rightarrow p_1,p_4\rightarrow p_3,p_3\rightarrow p_4)$.
\end{enumerate}
\paragraph{Soft Pentagon Integrals.}

We also need the soft limits of two configurations of the one-loop pentagon with one off-shell leg. 
\begin{enumerate}
\item Unlike the other one-loop soft integrals we require, the pentagon integral appearing in $\mathcal{M}^S_{13}$ does not permit a closed expression in terms of simple ${}_2F_1$ hypergeometric functions. Nonetheless, we are able to derive a compact Mellin-Barnes representation.
\beq\bsp
\label{eq:beastpenta}
\scalebox{0.4}{
   \begin{picture}(280,170) (0,70)
    \SetColor{Black}
    \SetWidth{4.0}
    \Line(192,32)(224,112)
    \Line(144,176)(64,112)
    \Line(64,112)(96,32)
    \Line(96,32)(192,32)
    \Line(224,112)(144,176)
    \Line[arrowpos=0.5,arrowlength=7.5,arrowwidth=4,arrowinset=0.2](144,208)(144,176)
    \Line[arrowpos=0.5,arrowlength=7.5,arrowwidth=4,arrowinset=0.2](272,128)(224,112)
    \Line[arrowpos=0.5,arrowlength=7.5,arrowwidth=4,arrowinset=0.2](208,0)(192,32)
    \Line[arrowpos=0.5,arrowlength=7.5,arrowwidth=4,arrowinset=0.2](32,128)(64,112)
    \Line[arrowpos=0.5,arrowlength=7.5,arrowwidth=4,arrowinset=0.2](80,0)(96,32)
    \SetWidth{8.0}
    \Line[arrowpos=0.5,arrowlength=7.5,arrowwidth=8,arrowinset=0.2](144,208)(144,176)
    \Text(10,128)[lb]{\huge{\Black{$p_1$}}}
    \Text(70,-18)[lb]{\huge{\Black{$p_3$}}}
    \Text(210,-18)[lb]{\huge{\Black{$p_4$}}}
    \Text(280,128)[lb]{\huge{\Black{$p_2$}}}
    \Text(110,220)[lb]{\huge{\Black{$-p_{1234}$}}}
  \end{picture}} \nonumber &= \int\frac{d^Dk}{i\pi^{D/2}} \frac{1}{k^2(k+p_3)^2(k-p_4)^2(2kp_2 -2p_2 p_4)(-2p_1 p_3-2kp_1)}\\
   &\nonumber\\
   \nonumber=\frac{2 \epsilon  (2 \epsilon +1) (-s_{34})^{-\epsilon -2}}{ s_{12} \Gamma (1-2 \epsilon )} &\int_{-i\infty}^{+i\infty}\frac{\d z_1\,\d z_2\,\d z_3}{(2\pi i)^3}\\
   \nonumber \times  \left(\frac{s_{13}+s_{14}}{s_{13}}\right)^{z_3}\,& \left(\frac{s_{24}}{s_{23}+s_{24}}\right)^{z_2} \left(\frac{s_{13} (s_{23}+s_{24})}{s_{12} s_{34}}\right)^{z_1} \\
   \times  \Gamma (z_1+1) \Gamma (-z_2&) \Gamma (-z_3) \Gamma (z_2-z_1) \Gamma (z_3-z_1) \Gamma (z_1+\epsilon +2) \\
   \nonumber \times  \Gamma (-z_2-\epsilon -1) \,&\Gamma (z_1 +z_2-z_3+1) \Gamma (-z_1+z_3-\epsilon -1) .
\esp\eeq
\item The pentagon integral required for $\mathcal{M}^S_{14}$ is given by
\beq\bsp
\scalebox{0.4}{
   \begin{picture}(280,170) (0,70)
    \SetColor{Black}
    \SetWidth{4.0}
    \Line(192,32)(224,112)
    \Line(144,176)(64,112)
    \Line(64,112)(96,32)
    \Line(96,32)(192,32)
    \Line(224,112)(144,176)
    \Line[arrowpos=0.5,arrowlength=7.5,arrowwidth=4,arrowinset=0.2](144,208)(144,176)
    \Line[arrowpos=0.5,arrowlength=7.5,arrowwidth=4,arrowinset=0.2](272,128)(224,112)
    \Line[arrowpos=0.5,arrowlength=7.5,arrowwidth=4,arrowinset=0.2](208,0)(192,32)
    \Line[arrowpos=0.5,arrowlength=7.5,arrowwidth=4,arrowinset=0.2](32,128)(64,112)
    \Line[arrowpos=0.5,arrowlength=7.5,arrowwidth=4,arrowinset=0.2](80,0)(96,32)
    \SetWidth{8.0}
    \Line[arrowpos=0.5,arrowlength=7.5,arrowwidth=8,arrowinset=0.2](144,208)(144,176)
    \Text(10,128)[lb]{\huge{\Black{$p_1$}}}
    \Text(70,-18)[lb]{\huge{\Black{$p_4$}}}
    \Text(210,-18)[lb]{\huge{\Black{$p_2$}}}
    \Text(280,128)[lb]{\huge{\Black{$p_3$}}}
    \Text(110,220)[lb]{\huge{\Black{$-p_{1234}$}}}
  \end{picture}} \nonumber &= \int\frac{d^Dk}{i\pi^{D/2}} \frac{1}{k^2(-2p_1p_4-2 k p_1)(2kp_2-2p_2p_3)(2kp_2 )(k+p_4)^2}\\
   &\nonumber\\
   =s_{12}^{\epsilon } (-s_{14})^{-\epsilon -1} &(-s_{23})^{-\epsilon -2}\frac{ \Gamma (1-\epsilon ) \Gamma (\epsilon +1)^2}{\epsilon ^2} \, _2F_1\left(1,\epsilon +1;1-\epsilon ;-\frac{s_{24}}{s_{23}}\right)\\
   \nonumber+s_{12}^{\epsilon } (-s_{14})^{-\epsilon -1}(& -s_{24})^{-\epsilon -1} \frac{2 \Gamma (1-\epsilon )^3 \Gamma (\epsilon +1)^2}{s_{23} \epsilon ^2 \Gamma (1-2 \epsilon )}.
\esp\eeq

\end{enumerate}

\subsection{The collinear region}
In this section we discuss the reduction to master integrals in the collinear region. Unlike the soft and the hard regions discussed in the previous sections, the phase-space integrals in the collinear region cannot be reduced to master integrals using standard techniques.
This is due to the appearance of propagators which are non-linear in the combined phase space and loop Mandelstam invariants. Effectively, the IBP relations do not close on certain topologies, and we are not aware of a method capable of dealing with IBP identities relating integral across different topologies.
However, as we already pointed out in Section~\ref{sec:SoftExpansion}, all the one-loop integrals appearing in the collinear region have the property that their denominators only depend on the loop momentum $k$ through the Lorentz invariants $k^2,k\cdot p_1$ and $k\cdot p_2$. This brings as a consequence that there exist relations
among the denominators, which can be used to find partial fraction identities to reduce all pentagons and boxes to a number of different triangle topologies. After integration over the loop momentum these triangles can be identified as linear combinations of only two types of bubbles.
As a consequence, we can write the contribution from the collinear region in the form
\beq\label{eq:collinear_bubbles}
C^{(1,c)}_{ij\to klH}(z) =\int \mathrm d \Phi_3^S\,  \bigg[A(\{s_{ij}\},\eps,\bar z)\mathrm{Bub}(s_{13})
+B(\{s_{ij}\},\eps,\bar z)\mathrm{Bub}(s_{13}+s_{14}) \bigg]\,,
\eeq
where $A$ and $B$  are expanded as power series in $\bar{z}$, whose coefficients are rational functions of the Mandelstam invariants and the dimensional regulator $\eps$. We dropped the real part of the loop integrals, because they are real for $s_{13}$, $s_{14}<0$. We emphasise that eq.~\eqref{eq:collinear_bubbles} is true to all orders in the expansion parameter $\bar z$ and can be proved simply by inspecting the propagators in the collinear region.

Next, we express bubble integrals in terms of tadpoles via the identity
\beq\label{eq:Bub_to_Tad}
\mathrm{Bub}(s)=-c_\Gamma\,\frac{1-\eps}{1-2\eps} \frac{1}{\Gamma(1+\eps)}
\frac{1}{s} \int \frac{\d^Dk}{i\pi^{D/2}} \frac{1}{k^2+s}\,.
\eeq
Since the single denominator in eq.~\eqref{eq:Bub_to_Tad} is clearly linear in all Lorentz invariants, this representation allows for a straightforward and efficient application of the IBP reduction technique to the combined soft phase space and (now trivial) one loop integral.
We find that in the collinear region we can reduce all integrals to just four \emph{soft} master integrals,
\beq\bsp
\mathcal M_1^C &\,=
\begin{picture}(130,28) (7,28)
    \SetWidth{1.0}
    \SetColor{Black}
    \SetWidth{2.0}
    \Arc[clock](47.582,-34.508)(83.634,93.141,49.408)
    \Line[double,sep=3](43,8)(102,29)
    \Arc(97.418,112.508)(83.634,-130.592,-86.859)
    \Line[dash,dashsize=3](82,59)(82,-1)
    \Line[arrow,arrowpos=0.5,arrowlength=6.667,arrowwidth=2.667,arrowinset=0.2](21,8)(43,8)
    \Text(14,5)[lb]{\small{\Black{$2$}}}
    \Line[arrow,arrowpos=0.5,arrowlength=6.667,arrowwidth=2.667,arrowinset=0.2](21,49)(43,49)
    \Text(14,45)[lb]{\small{\Black{$1$}}}
    \Text(124,37)[lb]{\small{\Black{$1$}}}
    \Text(124,15)[lb]{\small{\Black{$2$}}}
    \Line[arrow,arrowpos=0.5,arrowlength=6.667,arrowwidth=2.667,arrowinset=0.2](102,29)(122,39)
    \Line[arrow,arrowpos=0.5,arrowlength=6.667,arrowwidth=2.667,arrowinset=0.2](102,29)(122,19)
    \SetWidth{1.0}
    \Line[dash,dashsize=0.6](43,49)(43,8)
    \SetWidth{2.0}
    \Arc(82.5,28.5)(44.503,152.571,207.429)
    \Arc[clock](3.5,28.5)(44.503,27.429,-27.429)
  \end{picture}
 = \int \mathrm d\Phi_3^S\, (-s_{13}-s_{14})\,\mathrm{Bub}(s_{13}+s_{14}) \,,\\
\mathcal M_2^C&\,=\begin{picture}(130,65) (7,28)
    \SetWidth{1.0}
    \SetColor{Black}
    \SetWidth{2.0}
    \Arc[clock](90.5,-32.071)(62.145,139.849,79.336)
    \Line(43,49)(102,29)
    \Arc[double,sep=3](50.545,80.182)(72.575,-95.968,-44.848)
    \Line[dash,dashsize=3](82,59)(82,-1)
    \Line[arrow,arrowpos=0.5,arrowlength=6.667,arrowwidth=2.667,arrowinset=0.2](21,8)(43,8)
    \Text(14,5)[lb]{\small{\Black{$2$}}}
    \Line[arrow,arrowpos=0.5,arrowlength=6.667,arrowwidth=2.667,arrowinset=0.2](21,49)(43,49)
    \Text(14,45)[lb]{\small{\Black{$1$}}}
    \Text(124,37)[lb]{\small{\Black{$1$}}}
    \Text(124,15)[lb]{\small{\Black{$2$}}}
    \Line[arrow,arrowpos=0.5,arrowlength=6.667,arrowwidth=2.667,arrowinset=0.2](102,29)(122,39)
    \Line[arrow,arrowpos=0.5,arrowlength=6.667,arrowwidth=2.667,arrowinset=0.2](102,29)(122,19)
    \SetWidth{1.0}
    \Line[dash,dashsize=0.6](43,49)(43,8)
    \SetWidth{2.0}
    \Arc(82.5,28.5)(44.503,152.571,207.429)
    \Arc[clock](3.5,28.5)(44.503,27.429,-27.429)
  \end{picture}
 = \int \mathrm d\Phi_3^S\, (-s_{13})\,\mathrm{Bub}(s_{13})\,, \\
\mathcal M_3^C &\,=\begin{picture}(130,65) (7,28)
    \SetWidth{2.0}
    \SetColor{Black}
    \Line(67,49)(102,8)
    \SetWidth{1.0}
    \SetColor{White}
    \Vertex(89,24){4.243}
    \SetColor{Black}
    \SetWidth{2.0}
    \Line(43,49)(102,49)
    \Line[dash,dashsize=3](82,59)(82,-1)
    \Line(102,49)(102,8)
    \Line[double,sep=3](43,8)(102,29)
    \Line[arrow,arrowpos=0.5,arrowlength=6.667,arrowwidth=2.667,arrowinset=0.2](21,8)(43,8)
    \Text(14,5)[lb]{\small{\Black{$2$}}}
    \Line[arrow,arrowpos=0.5,arrowlength=6.667,arrowwidth=2.667,arrowinset=0.2](21,49)(43,49)
    \Text(14,45)[lb]{\small{\Black{$1$}}}
    \SetWidth{1.0}
    \Line[dash,dashsize=0.6](43,49)(43,8)
    \SetWidth{2.0}
    \Arc(82.5,28.5)(44.503,152.571,207.429)
    \Arc[clock](3.5,28.5)(44.503,27.429,-27.429)
    \Text(124,5)[lb]{\small{\Black{$2$}}}
    \Text(124,45)[lb]{\small{\Black{$1$}}}
    \Line[arrow,arrowpos=0.5,arrowlength=6.667,arrowwidth=2.667,arrowinset=0.2](102,49)(122,49)
    \Line[arrow,arrowpos=0.5,arrowlength=6.667,arrowwidth=2.667,arrowinset=0.2](102,8)(122,8)
  \end{picture}
    = \int \mathrm d\Phi_3^S\, \frac{(-s_{13}-s_{14})}{s_{13}s_{24}s_{34}} \,\mathrm{Bub}(s_{13}+s_{14})\,, \\
\mathcal M_4^C &\,=
 \begin{picture}(130,65) (7,28)
    \SetWidth{2.0}
    \SetColor{Black}
    \Line(43,8)(84,35)
    \SetWidth{1.0}
    \SetColor{White}
    \Vertex(78,27){7}
    \SetColor{Black}
    \SetWidth{2.0}
    \Line(43,49)(102,29)
    \SetWidth{1.0}
    \Line[dash,dashsize=0.6](43,49)(43,29)
    \SetWidth{2.0}
    \Arc(50.5,39)(12.5,126.87,233.13)
    \Arc[clock](35.5,39)(12.5,53.13,-53.13)
    \Line(43,8)(43,29)
    \Line[dash,dashsize=3](69,59)(69,-1)
    \Line[arrow,arrowpos=0.5,arrowlength=6.667,arrowwidth=2.667,arrowinset=0.2](21,8)(43,8)
    \Text(14,5)[lb]{\small{\Black{$2$}}}
    \Line[arrow,arrowpos=0.5,arrowlength=6.667,arrowwidth=2.667,arrowinset=0.2](21,49)(43,49)
    \Text(14,45)[lb]{\small{\Black{$1$}}}
    \Text(124,37)[lb]{\small{\Black{$1$}}}
    \Text(124,15)[lb]{\small{\Black{$2$}}}
    \Line[arrow,arrowpos=0.5,arrowlength=6.667,arrowwidth=2.667,arrowinset=0.2](102,29)(122,39)
    \Line[arrow,arrowpos=0.5,arrowlength=6.667,arrowwidth=2.667,arrowinset=0.2](102,29)(122,19)
    \Line[double,sep=3](43,29)(105,29)
  \end{picture}
  = \int \mathrm d\Phi_3^S \,\frac{(-s_{13})}{s_{24}s_{34}}\, \mathrm{Bub}(s_{13}) \,,
\esp\eeq
\vskip 0.7cm
\noindent
where the dotted lines represent numerator factors. We stress that the master integrals appearing in the collinear region are again soft integrals, and hence they can be computed using the same techniques as the master integrals in the soft and hard regions. This will be discussed in Section~\ref{sec:master_computation}.


\section{Evaluation of soft one-loop integrals}
\label{sec:master_computation}
In the previous section we have seen that, independently of the region they originate from, the master integrals appearing in the threshold expansion are always soft integrals, i.e., they all take the form
\beq\label{eq:soft_integral_prototype}
\cM_i^I = \int \d\Phi_3^S\,F_i^I(p_1,p_2,p_3,p_4;\epsilon)\,,
\eeq
where $\d\Phi_3^S$ denotes the soft phase-space measure of eq.~\eqref{eq:soft_PS_measure}, and the integrand is homogeneous with respect to a rescaling of the soft momenta as well as of the initial-state momenta $p_1$ and $p_2$,
\beq\bsp\label{eq:integrand_scaling}
F_i^I(\lambda\,p_1,p_2,p_3,p_4;\epsilon) &\,= \lambda^{\alpha_{iI}}\,F_i^I(p_1,p_2,p_3,p_4;\epsilon)\,,\\
F_i^I(p_1,\lambda\,p_2,p_3,p_4;\epsilon) &\,= \lambda^{\beta_{iI}}\,F_i^I(p_1,p_2,p_3,p_4;\epsilon)\,,\\
F_i^I(p_1,p_2,\lambda\,p_3,\lambda\,p_4;\epsilon) &\,= \lambda^{\gamma_{iI}}\,F_i^I(p_1,p_2,p_3,p_4;\epsilon)\,,
\esp\eeq
for some $\alpha_{iI}$, $\beta_{iI}$, $\gamma_{iI}$.

In this section we present methods to evaluate soft integrals analytically. We discuss two methods: the first is based on the derivation of a Mellin-Barnes representation for soft integrals, while the second exploits a particular phase-space factorisation as well as the homogeneity of the soft integrals to derive a parametric integral representation for soft integrals. For an additional method, using differential equation techniques, we refer to ref.~\cite{Zhu:2014fma}. We note that although we concentrate exclusively on the case of double-real emissions at one loop, all these techniques can in principle be generalised to arbitrary numbers of loops and legs in an obvious way.

\subsection{Mellin-Barnes representations for soft integrals}
In this section we describe a general method to write a given soft integral as a (possibly multi-fold) Mellin-Barnes (MB) integral with poles at integer values of the integration variables. In ref.~\cite{triplereal} such a procedure was introduced for soft integrals of purely real emissions at tree-level. In this section we briefly review the procedure of ref.~\cite{triplereal}, and in the end we argue that it can easily be extended beyond tree-level. As the procedure is essentially identical to the purely real case, we will be brief and refer to ref.~\cite{triplereal} for details.

The procedure of ref.~\cite{triplereal} starts from the observation that at tree level one may assume without loss of generality that the integrand $F$ is a product of powers of two-particle Mandelstam invariants. Indeed, this can always be achieved by replacing every sum of two-particle invariants in the denominator by its MB representation, using the well-known formula
\beq
\frac{1}{(A+B)^\lambda} = \frac{1}{\Gamma(\lambda)}\,\int_{-i\infty}^{+i\infty}\frac{\d z}{2\pi i}\,\Gamma(-z)\,\Gamma(z+\lambda)\,\frac{A^z}{B^{z+\lambda}}\,,
\eeq
where the contour separates the poles at $z = n$ from those at $z = -\lambda - n$, $n \in \mathbb{N}$.

Next, we parametrise the soft phase space using the energies and the angles of the soft momenta in the center-of-mass frame of the initial-state system. We write, with $s_{12}=1$,
\beq
p_1= \frac{1}{2}(1,1,0,\ldots)\,, \quad p_2= \frac{1}{2}(1,-1,0,\ldots)\,, \quad p_i=\frac{1}{2}\,E_i\,\beta_i\,, \,i=3,4\,,
\eeq
where $\beta_i$ is the four-velocity in the direction of $p_i$. The soft phase-space measure becomes,
\beq
\d\Phi_3^S = (2\pi)^{3-2D}\,2^{2-2D}\,\delta(1-E_3-E_4)\,E_3^{D-3}\,E_4^{D-3}\,\d E_3\,\d E_4\,\d\Omega_3^{D-1}\,\d\Omega_4^{D-1}\,,
\eeq
where $\Omega_i^{D-1}$ parametrises the solid angle of the soft momentum $p_i$. Due to our assumption that the integrand is a product of powers of two-particle Mandelstam invariants, the integration over the energies is simply a Beta function.
The remaining angular integrals can be written as MB integrals with poles at most at integer values of the integration variables~\cite{vanNeerven:1985xr,Somogyi:2011ir}.

If we follow this procedure, every tree-level soft integral can be written as a multifold MB integral with poles at integer values of the integration variables. If the integrand contains loop integrals which evaluate to complicated special functions, this claim is no longer necessarily true. It does, however, stay true if the loop integral itself admits an MB representation of the same type. In our case, the integrand contains at most hypergeometric functions, which admit the MB representation
\beq\bsp
{}_2F_1(a,b;c;x) &\,= \frac{\Gamma(c)}{\Gamma(a)\Gamma(b)}\int_{-i\infty}^{+i\infty}\frac{\d z}{2\pi i}\,\Gamma(-z)\,\frac{\Gamma(z+a)\Gamma(z+b)}{\Gamma(z+c)}\,(-x)^z\,.
\esp\eeq
An exception is one pentagon integral, which we did not express in terms of simple hypergeometric functions, but it still admits a multifold MB representation of a similar type, see eq.~\eqref{eq:beastpenta}.
We can therefore obtain an MB representation of the desired type for all soft integrals considered in this paper.

The MB integrals obtained in this way can be evaluated using standard techniques. In some cases it is possible to perform all the MB integration in closed form without expanding in $\eps$, and one obtains generalised hypergeometric functions,
\beq
{}_pF_q(a_1,\ldots,a_p;b_1,\ldots,b_q;z) = \sum_{n=0}^\infty\frac{\prod_{i=1}^p(a_i)_n}{\prod_{j=1}^q(b_j)_n}\,\frac{z^n}{n!}\,.
\eeq
Whenever we are not able to obtain a closed expression in terms of hypergeometric functions, we resolve the poles in $\eps$ using standard techniques~\cite{MB}. The result is a Laurent series in $\eps$ whose coefficients are MB integrals whose contours are straight vertical line. In all cases these integrals can be evaluated numerically in a  fast and efficient way. Alternatively, one can close the integration contours at infinity and sum up the residues of the poles of the Gamma functions in terms of nested harmonic sums~\cite{Vermaseren:1998uu} that evaluate to multiple zeta values.


\subsection{Soft integrals from phase-space factorisation}
We present in this section an alternative way to compute the soft integrals of Section~\ref{sec:IBPandSoftMasters}, based on a factorisation of the phase space that separates the soft part of the phase space from the phase space for the emission of the Higgs boson.

We start by writing the phase space for the production of a Higgs bosons in association with two massless partons as a convolution
\beq
\d\Phi_3 = \int\frac{\d\mu^2}{2\pi}\,\d\Phi_2(m_H^2,\mu^2;p_{12})\,\d\Phi_2(0,0;Q)\,,
\eeq
where $\d\Phi_2(m_1^2,m_2^2;q)$ denotes the phase-space measure for the decay of a heavy state with momentum $q$ into two particles with masses $m_1$ and $m_2$,
\beq
\d\Phi_2(m_1^2,m_2^2;q) = (2\pi)^{D}\,\delta^{(D)}(q-q_1-q_2)\,\frac{\d^Dq_1}{(2\pi)^{D-1}}\,\delta_+(q_1^2-m_1^2)\,\frac{\d^Dq_2}{(2\pi)^{D-1}}\,\delta_+(q_2^2-m_2^2)\,.
\eeq
Note that for now we work with the full phase-space measure, and we expand in $\bar{z}$ at the end. A soft integral of the type~\eqref{eq:soft_integral_prototype} can then be written as
\beq\label{eq:M_as_F}
\cM_i^I = \int\frac{\d\mu^2}{2\pi}\,\d\Phi_2(m_H^2,\mu^2;p_{12})\,\cF_i^I(p_1,p_2,Q)\,,
\eeq
where we defined
\beq\label{eq:cF_def}
\cF_i^I(p_1,p_2,Q) = \int \d\Phi_2(0,0;Q)\,F_i^I(p_1,p_2,p_3,p_4;\epsilon)\,.
\eeq

Let us first concentrate on the computation of the integral $\cF_i^I$. The homogeneity of the integrand of the original soft integral, eq.~\eqref{eq:integrand_scaling}, combined with the fact that the integration measure does not depend on $p_1$ and $p_2$, implies that $\cF_i^I$ is homogeneous under a rescaling of any of its arguments,
\beq\bsp
\cF_i^I(\lambda\,p_1,p_2,Q) &\,= \lambda^{\alpha_{iI}}\,\cF_i^I(p_1,p_2,Q)\,,\\
\cF_i^I(p_1,\lambda\,p_2,Q) &\,= \lambda^{\beta_{iI}}\,\cF_i^I(p_1,p_2,Q)\,,\\
\cF_i^I(p_1,p_2,\lambda\,Q) &\,= \lambda^{\gamma_{iI}}\,\cF_i^I(p_1,p_2,Q)\,.
\esp\eeq
Lorentz invariance then implies that the non-trivial functional dependence of $\cF_i^I$ can only be through the `cross ratio'
\beq
u = \frac{(p_1\cdot p_2)\,Q^2}{2\,(p_1\cdot Q)\,(p_2\cdot Q)}\,.
\eeq
Without loss of generality, we may write
\beq\label{eq:conformal_f}
\cF_i^I(p_1,p_2,Q) = \frac{(Q^2)^{(\gamma_{iI}-\alpha_{iI}-\beta_{iI})/2}}{(p_1\cdot Q)^{-\alpha_{iI}}\,(p_2\cdot Q)^{-\beta_{iI}}}\,f_i^I(u)\,.
\eeq
We can also give a kinematical meaning to the cross ratio $u$: it is related to the angle $\theta_{12}$ between $p_1$ and $p_2$ in the rest frame of $Q$,
\beq
u = \frac{1-\cos\theta_{12}}{2}\,.
\eeq
Note that this cross ratio is precisely the argument of the hypergeometric function in the box integral appearing in $\cMSTen$ and $\cMSTwentyThree$.

This suggests that the most natural frame in which to parametrise the phase space in eq.~\eqref{eq:cF_def} is the rest frame of $Q$. Writing $p_i = \frac{1}{2}E_i\,\beta_i$ with $\beta_i = (1,\vec n_i)$, the phase-space measure in this frame becomes
\beq
\d\Phi_2(0,0;Q) = 2^{1-D}\,(2\pi)^{2-D}\,(Q^2)^{(D-4)/2}\,\d\Omega_3^{D-1}\,,
\eeq
and the invariants are
\beq\bsp
s_{34} = Q^2\,, \qquad & s_{12} = 2\,\frac{(p_1\cdot Q)(p_2\cdot Q)}{Q^2}\,(1-\cos\theta_{12})\,,\\
s_{13} = -(p_1\cdot Q)\,\beta_1\cdot \beta_3\,,\qquad &s_{14} =  -(p_1\cdot Q)\,\overline{\beta}_1\cdot \beta_3\,,\\
s_{23} = -(p_2\cdot Q)\,\beta_2\cdot \beta_3\,,\qquad &s_{24} =  -(p_2\cdot Q)\,\overline{\beta}_2\cdot \beta_3\,,
\esp\eeq
with $\bar{\beta}_i = (1,-\vec n_i)$.
Here we used the fact that in the rest frame of $Q$ the final-state massless particles are back-to-back and their energy is $\sqrt{Q^2}/2$. The remaining integral over the solid angle can in all cases be performed using the formula~\cite{vanNeerven:1985xr,Somogyi:2011ir}
\beq\bsp\label{eq:angular_VN}
\int&\frac{\d\Omega_3^{D-1}}{(\beta_i\cdot\beta_3)^m\,(\beta_j\cdot\beta_3)^n} \\
&\,= 2^{2-m-n-2\eps}\,\pi^{1-\eps}\,\frac{\Gamma(1-m-\eps)\Gamma(1-n-\eps)}{\Gamma(1-\eps)\Gamma(2-m-n-2\eps)}\,{}_2F_1\left(m,n;1-\eps;1-\frac{\beta_i\cdot\beta_j}{2}\right)\,.
\esp\eeq
The scalar product appearing in the right hand side can take the following values,
\beq\bsp
\beta_1\cdot \beta_2 &\,= \overline{\beta}_1\cdot \overline{\beta}_2= 1-\cos\theta_{12} = 2u\,,\\
\beta_1\cdot \overline{\beta}_2 &\,= \overline{\beta}_1\cdot \beta_2= 1+\cos\theta_{12} = 2(1-u)\,.
\esp\eeq
Note that in some cases the integrand may contain hypergeometric functions depending on angular variables. We can reduce the problem to the angular integral~\eqref{eq:angular_VN} by introducing an MB representation for these hypergeometric functions. Eventually, we can in this way explicitly determine the function $f_i^I(u)$ in eq.~\eqref{eq:conformal_f}. We stress that although the computation was done in the rest frame of $Q$, the result is independent of the frame.

Inserting eq.~\eqref{eq:conformal_f} into eq.~\eqref{eq:M_as_F}, we are left with the computation of the remaining phase-space integral in eq.~\eqref{eq:M_as_F}. We write $Q = \alpha\,p_1+\beta\,p_2+Q_{\bot}$, with $p_1\cdot Q_{\bot} = p_2\cdot Q_{\bot} = 0$ and we work in the center-of-mass frame of the collision, and
the on-shell conditions for $Q$ and the Higgs boson enforce $\alpha+\beta=1+\ord(\bar{z})$ and $Q_{\bot}^2=\alpha(1-\alpha)-\mu^2$. Note that this implies $0<\alpha<1$ and $0<\mu^2<\alpha(1-\alpha)$. In terms of this parametrisation the invariants are
\beq
s_{12}=1\,,\qquad Q^2 = \mu^2 = \alpha\,(1-\alpha)\,u \,, \qquad 2\,p_1\cdot Q = 1-\alpha\,, \qquad 2\,p_2\cdot Q = \alpha\,,
\eeq
and the phase-space measure can be written as
\beq
\d\Phi_2(m_H^2,\mu^2;p_{12}) = \frac{1}{4}\,(2\pi)^{-2+2\eps}\,[\alpha\,(1-\alpha)\,(1-u)]^{-\eps}\,\d\alpha\,\d\Omega_Q^{D-2} + \ord(\bar{z})\,,
\eeq
where $\Omega_Q^{D-2}$ parametrises the solid angle of $Q$ in the center-of-mass frame. Putting everything together we see that the integrals over $\alpha$ and over the solid angle are trivial, and so we get, with $\delta=(\gamma-\alpha-\beta)/2$,
\beq
\cM_i^I = \frac{1}{2^{\alpha+\beta}\,(4\pi)^{2-\eps}}\,\frac{\Gamma(2-\eps+\alpha+\delta)\Gamma(2-\eps+\beta+\delta)}{\Gamma(1-\eps)\Gamma(4-2\eps+\alpha+\beta+2\delta)}\,\int_0^1\d u\,u^\delta\,(1-u)^{-\eps}\,f_i^I(u)\,.
\eeq
We obtain in this way a representation for the soft integral as a simple integral over the function $f_i^I$. This last integral is usually easy to carry out. In many cases it can be performed using the Euler-integral representation of the generalised hypergeometric function,
\beq\bsp
{}_3F_2&(a_1,a_2,a_3;b_1,b_2;z) \\
&\,= \frac{\Gamma(b_2)}{\Gamma(a_3)\Gamma(b_2-a_3)}\int_0^1\d u\,u^{a_3-1}\,(1-u)^{b_2-a_3-1}\,{}_2F_1(a_1,a_2;b_1;u\,z)\,.
\esp\eeq
In those cases where the function $f_i^I$ is more complicated, we can either insert an MB representation for it, or alternatively perform the integral over $u$ order by order in $\eps$ in terms of harmonic polylogarithms~\cite{Remiddi:1999ew}.

\subsection{Analytic results for the master integrals}
In this section we present the analytic results for all the master integrals introduced in Section~\ref{sec:IBPandSoftMasters}. As only the real part of the master integrals enters the final result for the cross section, we only present result for the real part, which we normalise according to
\beq
\textrm{Re}\left(\cM_i^I\right) = c_\Gamma\,\cos(\pi \eps)\,\Phi_3^S(\eps)\,M_i^I\,,
\eeq
where $c_\Gamma$ is defined in eq.~\eqref{eq:cGamma} and $\Phi_3^S(\eps)$ denotes the soft phase-space volume,
\beq
\Phi_3^S(\eps) = \int \d\Phi_3^S = \frac{1}{2\,(4\pi)^{3-2\eps}}\,\frac{\Gamma(1-\eps)^2}{\Gamma(4-4\eps)}\,.
\eeq
We have computed all the master integrals of Section~\ref{sec:IBPandSoftMasters} using the two different approaches described in the previous section, and we found complete agreement between the approaches. With the exception of the pentagon integrals $\MSThirtySix$ and $\MSThirtySeven$ we present the results for the master integrals as a Laurent expansion in the dimensional regulator up to terms of transcendental weight six for the soft and the hard regions and up to transcendental weight five for the collinear region. For $\MSThirtySix$ and $\MSThirtySeven$ we were only able to obtain the Laurent expansion up to terms of weight three and five respectively, which is however sufficient to compute the Higgs boson cross section up to finite terms~\footnote{At least for the 37 first terms in the threshold expansion, see ref.~\cite{Anastasiou:2015ema}.}. Whenever we were able to do so, we also include results for the master integrals valid to all orders in $\eps$ in terms of generalised hypergeometric functions, which can easily be expanded in $\eps$ using the {\tt HypExp} package~\cite{Huber:2005yg}. Note that all integrals presented in this section satisfy recursion relations with respect to the space-time dimension~\cite{Tarasov:1996br}. We have checked that our results satisfy these dimensional recurrence relations.

\subsection{Analytic results for the hard region}
\begin{align}
M_1^H & =\frac{1}{\eps\,(1-2\eps)}\\ 
\nonumber&=\frac{1}{\eps}+2+4\eps+8\eps^2+16\eps^3+32\eps^4+64\eps^5+\ord(\eps^6)\,,\\
\nonumber&\\
M_2^H&=-\frac{4(3-4\eps)(1-4\eps)}{\eps^4}\,{}_3F_2(1,1,-\eps;1-\eps,1-2\eps;1)\\
\nonumber&=
-\frac{18}{\epsilon ^4}+\frac{96}{\epsilon ^3}+\frac{1}{\epsilon ^2}(-96+12 \zeta _2)
+\frac{1}{\epsilon }\,(-64 \zeta _2+60 \zeta _3)
+64 \zeta _2-320 \zeta _3+186 \zeta _4\\
\nonumber&+\eps\,\left(320 \zeta _3-992 \zeta _4+24 \zeta _2 \zeta _3+444 \zeta _5\right) 
+\epsilon ^2\,(992 \zeta_4-128 \zeta_2 \zeta_3-2368 \
\zeta_5+60 \zeta_3^2\\
\nonumber&+1111 \zeta_6)   + \ord(\eps^3)\,.  
\end{align}

\subsection{Analytic results for the soft region}
\begin{align}
\MSOne&=\frac{\Gamma (4-4 \epsilon ) \Gamma (1-2 \epsilon )^2 \Gamma (1+\epsilon)}{2 \epsilon ^4 \Gamma (2-6 \epsilon ) \Gamma (1-\epsilon )}\\
\nonumber&=\frac{3}{\epsilon^4}-\frac{4}{\epsilon^3}+\frac{1}{\epsilon^2}\left(24-18\zeta_2\right)
+\frac{1}{\epsilon}\left(112+24\zeta_2-138\zeta_3\right)
+672-144\zeta_2+184\zeta_3
-621\zeta_4\\
\nonumber&
+\left(4032-672\zeta_2-1104\zeta_3+828\zeta_4+828\zeta_2\zeta_3-4014\zeta_5\right)\epsilon
+\Big(24192-4032\zeta_2-5152\zeta_3\\
\nonumber&-4968\zeta_4+5352\zeta_5-1104\zeta_2\zeta_3+3174\zeta_3^2-\frac{27501}{2}\zeta_6\Big)\epsilon^2+\ord(\eps^3)\,,
\end{align}
%
\begin{align}
\MSTwo&= \frac{\Gamma (4-4 \epsilon ) \Gamma (1-2 \epsilon )^3 \Gamma (1+\epsilon )}{\epsilon ^2 \Gamma (4-6 \epsilon ) \Gamma (1-\epsilon )^3}\\
\nonumber&=\frac{1}{\epsilon^2}+\frac{11}{3\epsilon}+\
\frac{61}{3}-5\zeta_2
+\Big(117-\frac{55}{3}\zeta_2-44 \zeta_3\Big)\epsilon
+\Big(687-\frac{305}{3}\zeta_2-\frac{484}{3}\zeta_3-\frac{869}{4}\zeta_4\Big)\epsilon^2\\
\nonumber&+\Big(4077-585\zeta_2-\frac{2684}{3}\zeta_3-\frac{9559}{12}\zeta_4+220\zeta_2\zeta_3-1332\zeta_5\Big)\epsilon^3
+\Big(24327-3435\zeta_2\\
\nonumber&-5148\zeta_3-\frac{53009}{12}\zeta_4+\frac{2420}{3}\zeta_2\zeta_3-4884\zeta_5+968\zeta_3^2-\frac{79655}{16}\zeta_6\Big)\epsilon^4+\ord(\eps^5)\,,\\
\nonumber&\\
\MSFour&= \frac{\Gamma (4-4 \epsilon ) \Gamma (2-3 \epsilon )}{(1-2 \epsilon )^2 \epsilon  \Gamma (4-6 \epsilon ) \Gamma (1-\epsilon )}\\
\nonumber&=\frac{1}{\epsilon}+\frac{14}{3}
+(24-6\zeta_2)\epsilon
+\Big(\frac{400}{3}-28\zeta_2-42\zeta_3\Big)\epsilon^2
+\Big(\frac{2320}{3}-144\zeta_2-196\zeta_3-195\zeta_4\Big)\epsilon^3\\
\nonumber&+\Big(4576-800\zeta_2-1008\zeta_3-910\zeta_4+252\zeta_2\zeta_3-1302\zeta_5\Big)\epsilon^4
+\Big(\frac{81920}{3}-4640\zeta_2-5600\zeta_3\\
\nonumber&-4680\zeta_4+1176\zeta_2\zeta_3-6076\zeta_5+882\zeta_3^2-\frac{9219}{2}\zeta_6\Big)\epsilon^5+\ord(\eps^6)\,,\\
\nonumber&\\
\MSSix&=
-\frac{10}{\epsilon^5}+\frac{220}{3\epsilon^4}+\frac{1}{\epsilon^3}\left(-160+96\zeta_2\right)
+\frac{1}{\epsilon^2}\Big(\frac{320}{3}-704\zeta_2+672\zeta_3\Big)
+\frac{1}{\epsilon}\Big(1536\zeta_2\\
\nonumber&-4928\zeta_3+2436\zeta_4\Big)
-1024\zeta_2+10752\zeta_3-17864\zeta_4-5760\zeta_2\zeta_3+16872\zeta_5\\
\nonumber&+\eps\,\left(-7168\zeta_3+38976\zeta_4+42240\zeta_2\zeta_3-123728\zeta_5-20160\zeta_3^2+34710\zeta_6\right)+\ord(\eps^2)\,,\\
\nonumber&\\
\MSTen &= -\frac{4 \Gamma (4-4 \epsilon ) \Gamma (1-3 \epsilon ) }{\epsilon  (1+\epsilon ) (1-2 \epsilon ) \Gamma (3-6 \epsilon ) \Gamma (1-\epsilon )} \, _3F_2(1,1,1-\epsilon ;2-3 \epsilon ,2+\epsilon;1)\\
\nonumber&=-\frac{12}{\epsilon}\zeta_2
-8\zeta_2-36\zeta_3
+(-112\zeta_2-24\zeta_3+33\zeta_4)\epsilon
+(-672\zeta_2-336\zeta_3+22\zeta_4+720\zeta_2\zeta_3\\
\nonumber&-450\zeta_5)\epsilon^2
+\Big(-4032\zeta_2-2016\zeta_3+308\zeta_4+480\zeta_2\zeta_3-300\zeta_5+1512\zeta_3^2+\frac{16881}{4}\zeta_6\Big)\epsilon^3\\
\nonumber& + \ord(\eps^4)\,.\\
\nonumber&\\
\MSTwelve&=-\frac{\Gamma (4-4 \epsilon ) \Gamma (1-2 \epsilon )^3 \Gamma (1+\epsilon)}{2 \epsilon ^3 (1+2 \epsilon) \Gamma (1-6 \epsilon ) \Gamma (1-\epsilon )^2 \Gamma (2-\epsilon )}\\
\nonumber&=
-\frac{3}{\epsilon^3}+\frac{25}{\epsilon^2}+\frac{1}{\epsilon}\left(-79+15\zeta_2\right)
+161-125\zeta_2+132\zeta_3
+\Big(-319+395\zeta_2-1100\zeta_3\\
\nonumber&+\frac{2607}{4}\zeta_4\Big)\epsilon
+\Big(641-805\zeta_2+3476\zeta_3-\frac{21725}{4}\zeta_4-660\zeta_2\zeta_3+3996\zeta_5\Big)\epsilon^2
+\Big(-1279\\
\nonumber&+1595\zeta_2-7084\zeta_3+\frac{68651}{4}\zeta_4+5500\zeta_2\zeta_3-33300\zeta_5-2904\zeta_3^2+\frac{238965}{16}\zeta_6\Big)\epsilon^3
+\ord(\eps^4)\,,
\end{align}
%
\begin{align}
\MSFourteen&=\frac{\Gamma (4-4 \epsilon ) \Gamma (1-3 \epsilon ) }{\epsilon ^3 (1-2 \epsilon) \Gamma (2-6 \epsilon ) \Gamma (1-\epsilon )}\, _3F_2(1,-\epsilon ,\epsilon ;1-2 \epsilon ,1-\epsilon ;1)\\
\nonumber&=\frac{6}{\epsilon^3}+\frac{4}{\epsilon^2}+\frac{1}{\epsilon}(56-42\zeta_2)
+336 -28\zeta_2-288\zeta_3
+\Big(2016-392\zeta_2-192\zeta_3-\frac{2433}{2}\zeta_4\Big)\epsilon\\
\nonumber&+\Big(12096-2352\zeta_2-2688\zeta_3-811\zeta_4+1980\zeta_2\zeta_3-8262\zeta_5\Big)\epsilon^2
+\Big(72576-14112\zeta_2\\
\nonumber&-16128\zeta_3-11354\zeta_4+1320\zeta_2\zeta_3-5508\zeta_5+6804\zeta_3^2-\frac{204663}{8}\zeta_6\Big)\epsilon^3+\ord(\eps^4)\,,\\
\nonumber&\\
\MSFifteen&=-\frac{2 \Gamma (4-4 \epsilon ) \Gamma (1-2 \epsilon )^3 \Gamma (1+\epsilon ) }{\epsilon ^3 \Gamma (3-6 \epsilon ) \Gamma (1-\epsilon )^3}\, _3F_2(1,1-2 \epsilon ,\epsilon ;2-3 \epsilon ,1-\epsilon ;1)\\
\nonumber&=-\frac{6}{\epsilon^3}-\frac{16}{\epsilon^2}+\frac{1}{\epsilon}\left(-112+36\zeta_2\right)
-672+96\zeta_2+306\zeta_3
+(-4032+672\zeta_2+816\zeta_3\\
\nonumber&+1410\zeta_4)\epsilon
+(-24192+4032\zeta_2+5712\zeta_3+3760\zeta_4-1842\zeta_2\zeta_3+8757\zeta_5)\epsilon^2\\
\nonumber&+\Big(-145152+34272\zeta_3+24192\zeta_2+26320\zeta_4-4912\zeta_2\zeta_3+23352\zeta_5-7824\zeta_3^2\\
\nonumber&+\frac{57177}{2}\zeta_6\Big)\epsilon^3+\ord(\eps^4)\,,\\
\nonumber&\\
\MSNineteen&=
\frac{11}{\epsilon^5}-\frac{143}{3\epsilon^4}
+\frac{1}{\epsilon^3}\left(-165-59\zeta_2\right)+\frac{1}{\epsilon^2}\Big(\frac{4301}{3}+\frac{767}{3}\zeta_2-488\zeta_3\Big)
+\frac{1}{\epsilon}\Big(-5005\\
\nonumber&+885\zeta_2+\frac{6344}{3}\zeta_3-\frac{9291}{4}\zeta_4\Big)
+15015 -\frac{23069}{3}\zeta_2+7320\zeta_3+\frac{40261}{4}\zeta_4+2576\zeta_2\zeta_3\\
\nonumber&-14502\zeta_5
+\epsilon\,\Big(-45045+26845\zeta_2
-\frac{190808}{3}\zeta_3+\frac{139365}{4}\zeta_4-\frac{33488}{3}\zeta_2\zeta_3+62842\zeta_5\\
\nonumber&+10708\zeta_3^2-\frac{849273}{16}\zeta_6\Big)+\ord(\eps^2)\,,\\
\nonumber&\\
\MSTwentyThree & =-\frac{16}{\epsilon ^5}+\frac{256}{3 \epsilon ^4}+\frac{368}{3 \epsilon ^3}
+\frac{1}{\epsilon ^2}\Big(-\frac{5920}{3}-240 \zeta _2+320 \zeta _3\Big)
+\frac{1}{\epsilon }\Big(\frac{27088}{3}+1760 \zeta _2\\
\nonumber&-\frac{7904 }{3}\zeta _3+3144 \zeta _4\Big)
 -34368-1680 \zeta _2+4352 \zeta _3-16852 \zeta _4+2976 \zeta _2 \zeta _3+14704 \zeta _5\\
\nonumber&+\epsilon\,\Big(124944-26240 \zeta _2+\frac{98912 }{3}\zeta _3-23488 \zeta _4+640 \zeta _2 \zeta _3-\frac{275248 }{3}\zeta _5+2496 \zeta _3^2\\
\nonumber&+104650 \zeta _6\Big)
+\ord(\eps^2)  \,,\\
\nonumber&\\
\MSThirtyThree & = -\frac{19}{\epsilon^4}+\frac{646}{3\epsilon^3}+\frac{1}{\epsilon^2}\Big(-\frac{3496}{3}+114\zeta_2\Big)
+\frac{1}{\epsilon}\Big(4864-1292\zeta_2+762\zeta_3\Big) - 19456\\
\nonumber&+6992\zeta_2-8636\zeta_3+3381\zeta_4+\epsilon\Big(77824-29184\zeta_2+46736\zeta_3-38318\zeta_4-4572\zeta_2\zeta_3\\
\nonumber&+22830\zeta_5\Big)+\ord(\eps^2)\,,\\
\nonumber&\\
\MSThirtyFour & = -\frac{18}{\eps^4}+\frac{186}{\eps^3}+\frac{1}{\eps^2}\Big(-846+120\zeta_2\Big)+\frac{1}{\eps}\Big(2730-1240\zeta_2+816\zeta_3\Big)-8190\\
\nonumber&+5640\zeta_2-8432\zeta_3+3516\zeta_4+\eps\Big(24570-18200\zeta_2+38352\zeta_3-36332\zeta_4-5376\zeta_2\zeta_3\\
\nonumber&+23880\zeta_5\Big)+\ord(\eps^2)\,,
\end{align}
%
\begin{align}
\MSThirtySix & = -\frac{50}{\eps^5}+\frac{374}{3\eps^4}+\frac{1}{\eps^3}\Big(\frac{4502}{3}+458\zeta_2\Big)+\frac{1}{\eps^2}\Big(-8022-\frac{3866}{3}\zeta_2+3256\zeta_3\Big)\\
\nonumber&+\ord(\eps^{-1})\,,\\
\nonumber&\\
\MSThirtySeven & = \frac{4}{\eps^5}+\frac{2}{3\eps^4}+\frac{1}{\eps^3}\Big(-210-55\zeta_2\Big)+\frac{1}{\eps^2}\Big(\frac{2554}{3}+\frac{391}{3}\zeta_2-394\zeta_3\Big)\\
\nonumber&+\frac{1}{\eps}\Big(-794+1725\zeta_2 +\frac{2638}{3}\zeta_3-\frac{5721}{4}\zeta_4\Big)-6834-\frac{27445}{3}\zeta_2+12810\zeta_3+\frac{9887}{4}\zeta_4\\
\nonumber&-10032\zeta_5+\ord(\eps)\,.\\ \nonumber
\end{align}


\subsection{Analytic results in the collinear region}

\begin{align}
M_1^C & = \frac{\Gamma (4-4 \epsilon ) \Gamma (3-3 \epsilon ) \Gamma (1+2 \epsilon)}{\epsilon\,(1-2 \epsilon )^2   \Gamma (5-5 \epsilon ) \Gamma (1-\epsilon ) \Gamma (1+\epsilon )} \\
\nonumber& = \frac{1}{2 \epsilon }+\frac{31}{24}
+\epsilon\,\Big(\frac{1241}{288}+\frac{1}{2}\zeta _2\Big)
+\epsilon^2\,\Big(\frac{62215}{3456}+\frac{31}{24} \zeta _2-7 \zeta _3\Big)
+ \epsilon^3\,\Big(\frac{3525449}{41472}+\frac{1241}{288} \zeta _2\\
\nonumber&-\frac{217}{12} \zeta _3-\frac{269 }{8}\zeta _4\Big)
+ \epsilon^4\,\Big(\frac{209334151}{497664}+\frac{62215}{3456} \zeta _2-\frac{8687 }{144}\zeta _3-\frac{8339 }{96}\zeta _4-7 \zeta _3 \zeta _2-189 \zeta _5\Big)\\
\nonumber&+\ord(\eps^5)\,,\\
\nonumber&\\
M_2^C & = \frac{\Gamma (4-4 \epsilon ) \Gamma (1-2 \epsilon )^2 \Gamma (1+2 \epsilon)}{\epsilon \, \Gamma (5-5 \epsilon ) \Gamma (1-\epsilon )^2 \Gamma (1+\epsilon )}\\
\nonumber&= \frac{1}{4 \epsilon }+\frac{37}{48}+\frac{1679 \epsilon }{576}
+\Big(\frac{87193}{6912}-\frac{9 }{2}\zeta _3\Big) \epsilon ^2
+\Big(\frac{4874375}{82944}-\frac{111 }{8}\zeta _3-\frac{81}{4} \zeta _4\Big) \epsilon ^3\\
\nonumber&+\Big(\frac{282663625}{995328}-\frac{1679 }{32}\zeta _3-\frac{999 }{16}\zeta _4-\frac{207 }{2}\zeta _5\Big) \epsilon ^4
+\ord(\eps^5)\,,\\
\nonumber&\\
M_3^C& =- \frac{\Gamma (4-4 \epsilon ) \Gamma (1-3 \epsilon ) \Gamma (1+2 \epsilon)}{\epsilon ^4\, (1-2 \epsilon) \Gamma (1-5 \epsilon ) \Gamma (1-\epsilon ) \Gamma (1+\epsilon)}\,{}_3F_2(1,1,-\epsilon ;1-2 \epsilon ,1-\epsilon ;1)\\
\nonumber &= -\frac{9}{\epsilon ^4}+\frac{48}{\epsilon ^3}+\frac{1}{\epsilon ^2}(-48-3 \zeta _2)+\frac{1}{\epsilon }(16 \zeta _2+156 \zeta _3)
-16 \zeta _2-832 \zeta _3+\frac{2853 }{4}\zeta _4\\
\nonumber&+\epsilon\,(832 \zeta _3-3804 \zeta _4+84 \zeta _2 \zeta _3+3624 \zeta _5)
+\ord(\eps^2)\,,\\
\nonumber& \\
M_4^C & = \frac{\Gamma (4-4 \epsilon ) \Gamma (1-2 \epsilon )^2 \Gamma (1+2 \epsilon) }{2 \epsilon ^4\, \Gamma (2-5 \epsilon ) \Gamma (1-\epsilon )^2 \Gamma (1+\epsilon)}\,{}_3F_2(1,\epsilon -1,-\epsilon ;1-2 \epsilon ,1-\epsilon ;1)\\
\nonumber &= \frac{3}{\epsilon ^4}-\frac{4}{\epsilon ^3}+\frac{1}{\epsilon ^2}(18-3 \zeta _2)
+\frac{1}{\epsilon }(60+4 \zeta _2-72 \zeta _3)
+290-12 \zeta _2+96 \zeta _3-\frac{1245 }{4}\zeta _4\\
\nonumber&+\epsilon\,(1420-44 \zeta _2-396 \zeta _3+415 \zeta _4+54 \zeta _3 \zeta _2-1467 \zeta _5)
+\ord(\eps^2)\,.
\end{align}


\section{Conclusions}
\label{sec:conclusions}
This paper discussed the expansion around threshold of the one-loop corrections to the production of a heavy colorless state in association with two partons. We introduced techniques to compute the coefficients in the expansion, in principle to any desired order, and to express the result in terms of a small set of soft master integrals. These results are the missing pieces which went into the computation of the inclusive gluon-fusion Higgs production cross section as an expansion around threshold~\cite{Anastasiou:2015ema,Anastasiou:2014vaa,Anastasiou:2014lda}.

Our main tool to reduce the coefficients appearing in the threshold expansion to a linear combination of soft master integrals is reverse unitarity, which allows one to map phase-space integrals 
to cuts of Feynman integral. We perform this expansion separately for the phase-space
measure and for the interference diagrams, and we observe that the phase-space integrals always reduce to integrals against the soft phase-space measure. The one-loop matrix elements are expanded in the soft limit using the strategy of regions, and the relevant regions are identified with the regions where the loop momentum is either hard, soft or collinear to one of the initial-state momenta.
In each region, we combine the expanded interference diagrams with 
the corresponding phase-space measure and use IBP identities to 
reduce them to soft master integrals, independently of the region.

The soft master integrals themselves are evaluated using two different approaches.
The first method allows to derive a Mellin-Barnes representation for the soft integrals with poles at most at integer locations, provided that a similar Mellin-Barnes representation for the loop integration can be obtained. 
In our case most of the Mellin-Barnes integrals can be done in closed form to all orders in the dimensional regulator. In the remaining cases we were able to obtain a representation of the integral as a Laurent expansion in the dimensional regulator.
The second method  
builds upon a specific factorisation of the phase space  
and exploits the knowledge of the scaling behavior of the
integral with the external momenta to arrive at a one-fold parametric integral representation for the soft integral, which can be performed using modern integration techniques.

The results for the soft master integrals presented in this paper are sufficient to obtain, at least, the first 37 terms in the threshold expansion of the N$^3$LO gluon-fusion cross section, and conjecturally they provide the full set of boundary conditions to compute all the phase-space master integrals in general kinematics appearing in inclusive N$^3$LO cross sections for the production of a heavy colorless state. We therefore anticipate that the results of this paper will have a substantial impact on future results for hadron collider cross sections at N$^3$LO, e.g., Drell-Yan production or Higgs production via bottom-quark fusion.


\section*{Acknowledgements}
This research was supported by the Swiss National Science Foundation (SNF) under
contracts 200021-143781 and  200020-149517, the European Commission
through the ERC grants ``MathAm'', ``IterQCD'', ``HEPGAME'' (320651) and the FP7 Marie Curie Initial Training Network ``LHCPhenoNet'' (PITN-GA- 2010-264564), by the U.S. Department of Energy
under contract no. DE-AC02-07CH11359 and the ``Fonds National de la Recherche Scientifique'' (FNRS), Belgium.


\appendix

\section{Derivation of eq.~(\ref{eq:trans_int})}
\label{app:transverse}
Here we wish to derive the tensor reduction of the following generic integral
\beq
I^{\mu_1..\mu_n}(p_1,p_2)=\int\frac{\d^Dk}{i\pi^{D/2}}\frac{k_\perp^{\mu_1}k_\perp^{\mu_2}\ldots k_\perp^{\mu_n}}{F(k,p_1,p_2)}
\eeq
which was stated in eq.~(\ref{eq:trans_int}). A general tensor Ansatz would yield that the tensor integral $I^{\mu_1..\mu_n}(p_1,p_2)$ can 
be written as a linear combination of $\{g^{\mu\nu},p_1^\mu,p_2^\nu\}$.
But since 
\beq
k_\perp.p_1=0=k_\perp.p_2,
\eeq
we must also have that 
\beq
(p_i)_{\mu_k}I^{\mu_1..\mu_n}(p_1,p_2)=0, \qquad \text{for $i=1,2$ and $k=1,2,..,n$}
\eeq
and therefore $I^{\mu_1..\mu_n}(p_1,p_2)$ can only depend on $\{g^{\mu\nu},p_1^\mu,p_2^\nu\}$ through the transverse metric
\beq
g_\perp^{\mu\nu}=g^{\mu\nu}-\frac{p_1^\mu p_2^\nu+p_2^\mu p_1^\nu}{p_1.p_2}\; .
\eeq
Given that $I^{\mu_1..\mu_n}(p_1,p_2)$ is fully symmetric under any permutations of the Lorentz indices $\mu_1,..,\mu_n$, the tensor structure
is fully determined to be that of eq.~(\ref{eq:trans_int}),i.e.
\beq
\label{eq:trans_Cdef}
I^{\mu_1..\mu_n}(p_1,p_2)=\frac{1}{C(n)}\, g_\perp^{\mu_1..\mu_n}\, I(n,p_1,p_2)
\eeq
where $g_\perp^{\mu_1..\mu_n}$ is defined in eq.~(\ref{eq:trans_tensor}), $C(n)$ still has to be determined and 
\beq
I(n,p_1,p_2)= \int\frac{\d^Dk}{i\pi^{D/2}}\frac{(k_\perp^2)^{n/2}}{F(k,p_1,p_2)}\;.
\eeq
To determine the coefficient $C(n)$ we contract each of any $n/2$ of the $n$ Lorentz indices on both sides of 
eq.~(\ref{eq:trans_Cdef}) with 
any one of the remaining $n/2$ indices. Due to the symmetry it is irrelevant how this contraction is done and the result is
\beq
C(n)=\, g_\perp^{\mu_1..\mu_n} g_{\mu_1\mu_2}..g_{\mu_{n-1}\mu_n}
\eeq
Using 
\beq
g_\perp^{\mu_1..\mu_n}=g_\perp^{\mu_1\mu_2}g_\perp^{\mu_3..\mu_n} + \sum_{l\neq k=3}^n g_\perp^{\mu_l\mu_2}g_\perp^{\mu_k\mu_1} g_\perp^{\mu_3..\not{\mu_k}..\not{\mu_l}..\mu_n}
\eeq
we can explicitly carry out the contraction over $\mu_1$ and $\mu_2$ to get
\bea
C(n) &=& \big(  D_\perp g_\perp^{\mu_3..\mu_n} + 2\sum_{l>k=3}^n g_\perp^{\mu_l\mu_k}    g_\perp^{\mu_3..\not{\mu_k}..\not{\mu_l}..\mu_n}\big) g_{\mu_3\mu_4}..g_{\mu_{n-1}\mu_n} \\
&=&  \big(  D_\perp  + (n-2)\big)g_\perp^{\mu_3..\mu_n} g_{\mu_3\mu_4}..g_{\mu_{n-1}\mu_n} \\
&=&  \big(  D_\perp  + (n-2)\big)C(n-2) 
\eea
where $D_\perp=(g_\perp)^\mu_{\;\;\mu}=D-2$ and in the second line we used the identity
\beq
\frac{n}{2}\,g_\perp^{\mu_1..\mu_n}=\sum_{l>k=1}^n g_\perp^{\mu_l\mu_k}    g_\perp^{\mu_1..\not{\mu_k}..\not{\mu_l}..\mu_n}\;.
\eeq
Given that $C(2)=D_\perp$ it follows that
\beq
C(n)=\prod_{i=1}^{n/2} (D_\perp +2(i-1))
\eeq
which completes the derivation of eq.~(\ref{eq:trans_int}).


\end{document}